\documentclass[amsmath,amssymb]{revtex4}
\usepackage{graphicx}
\usepackage{amscd,amsmath,amsthm,amsfonts,amssymb}
\def\[{\begin{equation}}
\def\]{\end{equation}}

\begin{document}
\title{Pattern transformation in higher-order lumps of the Kadomtsev-Petviashvili I equation}
\author{Bo Yang and Jianke Yang}
\affiliation{Department of Mathematics and Statistics, University of Vermont, Burlington, VT 05405, U.S.A.}
\begin{abstract}
Pattern formation in higher-order lumps of the Kadomtsev-Petviashvili I equation at large time is analytically studied.
For a broad class of these higher-order lumps, we show that two types of solution patterns appear at large time. The first type of patterns comprise fundamental lumps arranged in triangular shapes, which are described analytically by root structures of the Yablonskii--Vorob'ev polynomials. As time evolves from large negative to large positive, this triangular pattern reverses itself along the $x$-direction. The second type of patterns comprise fundamental lumps arranged in non-triangular shapes in the outer region, which are described analytically by nonzero-root structures of the Wronskian--Hermit polynomials, together with possible fundamental lumps arranged in triangular shapes in the inner region, which are described analytically by root structures of the Yablonskii--Vorob'ev polynomials. When time evolves from large negative to large positive, the non-triangular pattern in the outer region switches its $x$ and $y$ directions, while the triangular pattern in the inner region, if it arises, reverses its direction along the $x$-axis. Our predicted patterns at large time are compared to true solutions, and excellent agreement is observed.
\end{abstract}
\maketitle

\section{Introduction}
The Kadomtsev-Petviashvili (KP) equation was derived as a two-dimensional generalization of the Korteweg-de Vries equation for the evolution of weakly nonlinear plasma waves and shallow water waves \cite{KP1970,Ablowitz1979}. In the water wave context, this equation reads \cite{Ablowitz1979}
\[
\left[2f_{t}+3ff_x+(\frac{1}{3}-T)f_{xxx}\right]_x+f_{yy}=0,
\]
where the spatial coordinate $x$ is relative to a certain moving frame, $f(x,y,t)$ represents the water surface elevation, and $T$ is a dimensionless surface tension parameter. If the surface tension is large, i.e., $T>1/3$, which corresponds to very thin sheets of water, this equation is called KP-I. In this case, rescaling variables by
\[
y=\frac{\hat{y}}{\sqrt{3(T-\frac{1}{3}})}, \quad t=-\frac{2\hat{t}}{T-\frac{1}{3}}, \quad f=-2(T-\frac{1}{3})u
\]
and dropping the hats, this equation becomes
\[\label{KPI}
\left( u_t+ 6 u u_x +u_{xxx} \right)_x - 3u_{yy} =0.
\]
Note that the KP-I equation also arises in other branches of physics, such as nonlinear optics \cite{Pelinovsky1995} and Bose-Einstein condensates \cite{Tsuchiya_BEC2008}.

The KP-I equation (\ref{KPI}) is solvable by the inverse scattering transform \cite{Zakharov_book, Ablowitz_book}. It admits stable fundamental lump solutions that are bounded rational functions decaying in all spatial directions \cite{Petviashvili1976,Manakov1977,Ablowitz_Satsuma1979}. These lumps are the counterparts of solitons in the Korteweg-de Vries equation. In the water wave context, these lumps physically correspond to dips on the water surface due to the negative sign in the $f$ scaling above. The KP-I equation also admits a broad class of rational solutions that describe the interactions of these lumps. If individual lumps have distinct asymptotic velocities, then they would pass through each other without change in velocities or phases \cite{Manakov1977,Ablowitz_Satsuma1979}. But if they have the same asymptotic velocities, they would undergo novel anomalous scattering, where the lumps would separate from each other in new spatial directions that are very different from their original incoming directions \cite{Peli93b,Ablowitz97,Ablowitz2000}. In this article, we are concerned with this latter type of solutions, which we will call higher-order lumps (they are also called multi-pole lumps in the literature \cite{Ablowitz97,Ablowitz2000}).

Analytical expressions of higher-order lumps have been derived by a wide variety of methods before \cite{Peli93a,Peli93b,Ablowitz97,Peli98,Ablowitz2000,Dubard2010,Dubard2013,Ma2015,Chen2016,ClarksonDowie2017,Gaillard2018,Chang2018}. Gorshkov, et al. \cite{Peli93b} reported a second-order lump solution that describes the interaction and anomalous scattering of two lumps. Ablowitz et al. \cite{Ablowitz97,Ablowitz2000} derived higher-order lumps by the inverse scattering transform and Darboux transformation, and reproduced the solution in \cite{Peli93b} as a special case. They also showed that when $|t| \to \infty$, these higher-order lumps generically split into a certain number of fundamental lumps, whose relative  spatial separations grow in proportion to $|t|^q$, where $\frac{1}{3}\le q\le \frac{1}{2}$. In addition, some new lump patterns such as squares at large time were reported. Pelinovsky and Stepanyants \cite{Peli93a} reported a class of higher-order lump solutions that are stationary in a moving frame. Pelinovsky \cite{Peli98} studied rational solutions of the KP hierarchy and linked them to the dynamics of the Calogero--Moser hierarchy (but his Wronskian-form solutions for KP-I were not made real-valued and thus were not physical solutions). Dubard et al. \cite{Dubard2010,Dubard2013} constructed a class of higher-order KP-I lump solutions from higher-order rogue waves of the nonlinear Schr\"odinger equation, and graphically showed that such second- and third-order lump solutions split into triangles of fundamental lumps when $|t| \to \infty$. Chen et al. \cite{Chen2016} considered a certain class of higher-order lump solutions, and graphically observed that these solutions evolve from a vertical line of fundamental lumps to a horizontal line of fundamental lumps in the $(x,y)$ plane when time goes from negative infinity to infinity. They also predicted the locations of fundamental lumps inside the solution complex at $t=0$ by roots of certain polynomial equations; but such polynomial equations were not justified. Clarkson and Dowie \cite{ClarksonDowie2017} derived a second-order lump solution which incorporates the ones in \cite{Dubard2010,Dubard2013,Peli93b} as special cases. Gaillard \cite{Gaillard2018} studied a special class of higher-order lump solutions and reported lump patterns such as triangles and pentagons at $t=0$ when some internal parameters in such solutions get large. Chang \cite{Chang2018} studied the large-time asymptotics of higher-order lumps and showed that, for some special solutions, all lumps are located on a vertical line in the $(x,y)$ plane at large negative time but rotate to a horizontal line at large positive time. Ma \cite{Ma2015} derived a fundamental lump solution which contains more free parameters; but that solution can be made equivalent to the original lump solution as reported in \cite{Manakov1977,Ablowitz_Satsuma1979}. We note by passing that non-rational KP-I solutions in the form of a linear periodic chain of lumps, and those that describe the resonant collision between lumps and line solitons, have also been reported recently \cite{Zakharov2021,He_KPI_rogue_lump}.

In this article, we study pattern formation in higher-order lumps of the KP-I equation (\ref{KPI}). This work is motivated by our earlier work on pattern formation of rogue waves in various integrable systems \cite{YangYang21a, YangYang21b}, where we showed that universal rogue patterns appear when one of the internal parameters in rogue waves gets large, and those rogue patterns are analytically described by root structures of the Yablonskii--Vorob'ev polynomial hierarchy. For higher-order lumps of the KP-I equation, however, we will focus on their pattern formation at large time rather than at large parameters. In particular, we are interested to know how their patterns at large positive time relate to their patterns at large negative time. For a broad class of higher-order lump solutions, we will show that two types of lump patterns appear at large time. The first type of patterns comprise fundamental lumps arranged in triangular shapes, which are described analytically by root structures of the Yablonskii--Vorob'ev polynomials. As time evolves from large negative to large positive, this triangular pattern reverses itself along the $x$-direction. The second type of patterns comprise fundamental lumps arranged in non-triangular shapes in the outer region, which are described analytically by nonzero-root structures of the Wronskian--Hermit polynomials, together with possible fundamental lumps arranged in triangular shapes in the inner region, which are described analytically by root structures of the Yablonskii--Vorob'ev polynomials. When time evolves from large negative to large positive, the non-triangular pattern in the outer region switches its $x$ and $y$ directions, plus some rescaling along each direction, while the triangular pattern in the inner region, if it arises, reverses its direction along the $x$-axis. These dramatic pattern transformations with the elapse of time are fascinating. We have also compared these predicted patterns with true solutions, and excellent agreement is observed.

This paper is organized as follows. In Sec.~2, we present general higher-order lump solutions in the KP-I equation through Schur polynomials, and introduce Yablonskii--Vorob'ev and Wronskian--Hermit polynomials. In Sec.~3, we present our main analytical results on solution patterns at large time, and explain how these patterns transform from large negative time to large positive time. In Sec.~4, we illustrate our pattern predictions and compare them with true solutions. In Sec.~5, we provide proofs for our analytical results in Sec.~3. The last section summarizes our results, together with some discussions. In the Appendix, a brief derivation of our general higher-order lump solutions in Sec.~2 is given.

\section{Preliminaries}
The KP equation (\ref{KPI}) admits three important invariances. The first one is that it is invariant when $(x, t)\to (-x, -t)$. This invariance is important because it shows that KP-solution patterns are reversible in time (albeit with a sign switch in $x$). In earlier works \cite{Chen2016,Chang2018}, the authors showed that certain higher-order KP lumps evolve from a vertical line of fundamental lumps to a horizontal line of fundamental lumps in the $(x,y)$ plane when time goes from negative infinity to infinity. The above invariance indicates that a reverse pattern transformation could also occur, i.e., those higher-order KP lumps can also evolve from a horizontal line to a vertical line when time goes from negative infinity to infinity.

The second invariance of the KP equation (\ref{KPI}) is the Galilean invariance \cite{Weiss1985,Chen2016}, i.e., when
\[ \label{Galilean}
(x,y,t) \to (x+2ky+12k^2t, \, y+12kt, \, t),
\]
the KP solution $u(x, y, t)$ remains a solution. Here, $k$ is an arbitrary real constant. This invariance indicates that, if the overall solution complex has a $y$-direction velocity $12k$, then we can apply this invariance to remove that $y$-direction velocity. In doing so, the solution pattern in the $(x,y)$ plane would change as well through a linear transformation of shear type. This Galilean invariance is important, because it allows us to remove the overall $y$-direction velocity in a higher-order lump solution. More will be said on it later in this section.

The third invariance of the KP equation is scaling invariance, i.e., when
\[ \label{scaling}
(x, y, t, u)\to (\alpha x, \alpha ^2y, \alpha^3t, \alpha^{-2}u),
\]
the KP equation remains invariant. Here, $\alpha$ is any nonzero real constant. This invariance is useful since, when combined with the Galilean invariance above, it allows us to normalize the spectral parameter in the KP-lump solutions to be unity without any loss of generality. This we will do in Sec.~3.

\subsection{Explicit expressions of higher-order lumps}
In this paper, we consider pattern formation of higher-order lumps in the KP-I equation (\ref{KPI}).  General higher-order lump solutions have been derived by Ablowitz et al. \cite{Ablowitz2000} through Darboux transformation. Their solutions were given through determinants whose matrix elements involve differential operators with respect to the spectral parameter. For our analysis, those solution expressions are not explicit enough. Thus, we have derived these higher-order lumps again by the bilinear method. To present our solutions, we first introduce elementary Schur polynomials $S_k(\mbox{\boldmath $x$})$ with $ \emph{\textbf{x}}=\left( x_{1}, x_{2}, \ldots \right)$, which  are defined by the generating function
\begin{equation}\label{Elemgenefunc}
\sum_{n=0}^{\infty}S_n(\mbox{\boldmath $x$}) \epsilon^n
=\exp\left(\sum_{n=1}^{\infty}x_n \epsilon^n\right).
\end{equation}
More explicitly,
\begin{equation*}
S_{0} (\emph{\textbf{x}})=1,\quad  S_1(\mbox{\boldmath $x$})=x_1,
\quad S_2(\mbox{\boldmath $x$})=\frac{1}{2}x_1^2+x_2, \quad \cdots, \quad
S_{n}(\mbox{\boldmath $x$}) =\sum_{l_{1}+2l_{2}+\cdots+ml_{m}=n} \left( \ \prod _{j=1}^{m} \frac{x_{j}^{l_{j}}}{l_{j}!}\right).
\end{equation*}
Under these notations, our general higher-order KP-I lumps are given by the following theorem.
\begin{quote}
\textbf{Theorem 1} \hspace{0.05cm} \emph{General higher-order lumps of the KP-I equation (\ref{KPI}) are }
\begin{eqnarray}
  && u_{\Lambda} (x,y,t)=2 \partial_{x}^2 \ln \sigma, \label{Schpolysolu}
\end{eqnarray}
\emph{where }
\[\label{Blockmatrix}
\sigma(x,y,t)=\det_{1 \leq i,j \leq N}\left(m_{ij}\right),
\]
\[ \label{Schmatrimnij}
m_{i, j}=\sum_{\nu=0}^{\min(n_i,n_j)} \left[ \frac{|p|^2}{(p+p^*)^2}  \right]^{\nu} \hspace{0.06cm} S_{n_i-\nu}(\textbf{\emph{x}}^{+} +\nu \textbf{\emph{s}}+\textbf{\emph{a}}_{i} )  \hspace{0.06cm} S_{n_j-\nu}[(\textbf{\emph{x}}^{+})^* + \nu \textbf{\emph{s}}^*+ \textbf{\emph{a}}_{j}^*],
\]
\emph{$N$ is an arbitrary positive integer, $\Lambda \equiv (n_{1}, n_{2}, \cdots n_N)$ is a vector of arbitrary positive integers,}
\emph{$p$ is an arbitrary non-imaginary complex number, the asterisk `*' represents complex conjugation, the vector $\textbf{\emph{x}}^{+}=\left( x_{1}^{+}, x_{2}^{+},\cdots \right)$ is defined by }
\begin{eqnarray}
x_{k}^{+}= p \frac{1}{k!} x + p^2 \frac{2^k}{k!} \textrm{i} y +  p^3 \frac{3^k}{k!} (-4) t,  \label{defxrp}
\end{eqnarray}
\emph{the vector $\textbf{\emph{s}}=(s_1, s_2, \cdots)$ is defined through the expansion}
\begin{eqnarray}
\ln \left[\frac{1}{\kappa} \left(p+p^* \right) \left( \frac{e^{\kappa}-1}{p \hspace{0.05cm} e^{\kappa}+p^*} \right)  \right] = \sum_{j=1}^{\infty}s_{j} \hspace{0.05cm} \kappa^j,  \label{schurcoeffsr}
\end{eqnarray}
\emph{vectors $\textbf{\emph{a}}_{i}$ are}
\[\label{multiparas}
\textbf{\emph{a}}_{i}=\left( a_{i,1}, a_{i,2}, \cdots , a_{i,n_i}\right),
\]
\emph{and $a_{i, j} \hspace{0.05cm} (1\le i\le  N,  1\le j\le n_i)$ are free complex constants. }
\end{quote}

The proof of this theorem will be given in the Appendix.

\textbf{Remark 1.} In this theorem, positive integers $(n_{1}, n_{2}, \cdots n_N)$ do not have to be distinct if their corresponding vectors $\textbf{\emph{a}}_{i}$ are different. In such cases, by first rewriting the $\sigma$ determinant (\ref{Blockmatrix}) as a larger determinant as was done in Ref.~\cite{OhtaJY2012}, then linking Schur polynomials with different $\textbf{\emph{a}}_{i}$ vectors in that larger determinant by relations similar to Eq.~(167) in Ref.~\cite{YangYang3wave}, and finally applying row operations and parameter redefinitions to the resulting determinant, we can show that this $\sigma$ determinant (\ref{Blockmatrix}) with non-distinct integers $(n_{1}, n_{2}, \cdots n_N)$ can be reduced to one where the new integers $(\hat{n}_{1}, \hat{n}_{2}, \cdots \hat{n}_N)$ become distinct. Thus, in this paper, we will require positive integers $(n_{1}, n_{2}, \cdots n_N)$ to be distinct without loss of generality. In this case, we will also arrange them in the ascending order, i.e., $n_1<n_2<\cdots <n_N$.

\textbf{Remark 2.}
The higher-order lumps in Theorem 1 contain free complex parameters $p$ and $\textbf{\emph{a}}_{i}$ ($1\le i\le N$), totaling $1+n_1+n_2+\cdots n_N$. However, using techniques similar to that outlined in Remark 1, we can show that $N(N-1)/2$ of those parameters in $\{\textbf{\emph{a}}_{i}\}$ can be eliminated. Thus, the number of free complex parameters in these higher-order lumps can be reduced to $1+\rho$, where
\[ \label{defrho}
\rho=\sum_{i=1}^N n_i -\frac{N(N-1)}{2}.
\]
This number of free parameters matches that given in Ref.~\cite{Ablowitz2000} for solutions produced by Darboux transformation. In fact, from the derivation of Theorem 1 in the Appendix, we can see that our higher-order lumps given in this theorem by the bilinear method are equivalent to those derived in \cite{Ablowitz2000} by Darboux transformation, except that our expressions are more explicit.

\textbf{Remark 3.} The fundamental lump can be derived by taking $N=1$ and $n_1=1$ in Eq.~(\ref{Blockmatrix}). Through a shift of the $(x,y)$ axes, we can normalize $a_{1,1}=0$. Then, the resulting $\sigma_1(x,y,t)$ function can be reduced to
\begin{eqnarray}
&&\sigma_1=\left|x+2 \textrm{i}  p y  -12 p^2 t\right |^2  + \frac{1}{(p+p^*)^2}  \nonumber \\
&& \hspace{0.3cm} = \left[x-2p_i \hspace{0.03cm} y-12(p_r^2-p_i^2) \hspace{0.03cm} t \right]^2+\left[2p_r(y-12p_i \hspace{0.03cm} t) \right]^2+ \frac{1}{4p_r^2},      \label{Fundsigma}
\end{eqnarray}
where $p_r$ and $p_i$ are the real and imaginary parts of the spectral parameter $p$. The corresponding solution $u_1(x,y,t)$ through Eq.~(\ref{Schpolysolu}) moves at $x$-direction velocity of $12|p|^2$ and $y$-direction velocity of $12p_i$. By applying the Galilean invariance (\ref{Galilean}) with $k=p_i$, we can remove the $y$-direction velocity $12p_i$ and reduce $\sigma_1(x,y,t)$ to
\begin{eqnarray}
\sigma_1= \left(x-12p_r^2  t  \right)^2+\left(2p_r y \right)^2+ \frac{1}{4p_r^2}.
\end{eqnarray}
This means that, under Galilean invariance, we can take $p$ in the original fundamental lump to be purely real without loss of generality. Then, by utilizing the scaling invariance (\ref{scaling}) with $\alpha=p_r$, we can further normalize $p_r$ in the above $\sigma_1$ to be unity. The final simplified fundamental-lump expression is
\[ \label{defu1}
u_1(x,y,t)=2 \partial_{x}^2 \ln \left[\left(x-12t  \right)^2+4y^2+ \frac{1}{4}\right].
\]
This is a moving single lump with peak amplitude $16$, which is attained at the spatial location of $(x, y)=(12t, \hspace{0.05cm} 0)$.

\textbf{Remark 4.} In the general higher-order lump of Theorem 1, the whole solution complex moves at $x$-direction velocity $12|p|^2$ and $y$-direction velocity $12p_i$, plus some possible slower motion relative to that moving frame. In this general case, we can also use the Galilean invariance (\ref{Galilean}) to remove the $y$-direction velocity $12p_i$ of the complex, i.e., $p_i$ can be made to be zero. In addition, we can use the scaling invariance (\ref{scaling}) to normalize $p_r$ to unity. Thus, without any loss of generality, we can choose $p$ in the higher-order lump solution of Theorem 1 to be equal to one. For this reason, we will set $p=1$ in the remainder of this paper.

\textbf{Remark 5.} In \cite{Peli93a}, a class of higher-order lump solutions that are stationary in a moving frame was reported. Those special solutions satisfy the Boussinesq equation. Thus, they are special cases of Boussinesq rogue waves \cite{YangYangBoussi}. Those stationary higher-order lumps are also special cases of our solutions in Theorem 1 when the index vector $(n_{1}, n_{2}, \cdots n_N)$ and internal parameters $\{\textbf{\emph{a}}_{i}\}$ are properly chosen. Indeed, rational solutions in Theorem 1 would be stationary if the $\sigma$ function in (\ref{Blockmatrix}) satisfies the dimension-reduction condition $\sigma_t-V\sigma_x=0$, where $V$ is the velocity of the moving frame along the $x$-direction. In the bilinear derivation of Boussinesq rogue waves \cite{YangYangBoussi}, one needs to solve the bilinear $\tau$ equation of KP-I, together with this $\tau$'s dimension reduction condition $\tau_{x_3}-3\tau_{x_1}=C\tau$, where $x_1$ is proportional to $x$, $x_3$ proportional to $t$, and $C$ is a constant. Since this $\tau$ function turns out to be equal to $\sigma$ multiplying an exponential of a linear function of $x$ and $t$ \cite{YangYangBoussi}, we see that $\tau$'s dimension reduction condition is equivalent to $\sigma$'s dimension reduction condition after proper variable scalings. This means that constraints from $\tau$'s  dimension-reduction condition can be borrowed over and imposed on solutions of Theorem 1 in order to obtain stationary higher-order KP lumps. One of such constraints is on the index vector $(n_{1}, n_{2}, \cdots n_N)$, where $n_i=2i-1$ must be chosen \cite{YangYangBoussi}. In addition, internal parameters $\{\textbf{\emph{a}}_{i}\}$ also need to be constrained. For a different choice of differential operators than those in Eq.~(\ref{AiBj}) of the Appendix, this parameter constraint was derived in \cite{YangYangBoussi}. For the present choice of differential operators in Eq.~(\ref{AiBj}), this parameter constraint would be more complex. In this case, such a parameter constraint was worked out in \cite{ChenJunChao2018} for another integrable system under a different parameterization.

\subsection{Yablonskii--Vorob'ev polynomials and Wronskian-Hermit polynomials}
We will show in later text that patterns of certain higher-order lump solutions at large time are described by root structures of the Yablonskii--Vorob'ev polynomials and Wronskian-Hermit polynomials. Thus, these polynomials and their root structures will be introduced first.

\subsubsection{Yablonskii--Vorob'ev polynomials and their root structures}
Yablonskii-Vorob'ev polynomials arose in rational solutions of the second Painlev\'{e} equation ($\mbox{P}_{\mbox{\scriptsize II}}$) \cite{Yablonskii1959,Vorobev1965}. Later, a determinant expression for these polynomials was found in \cite{Kajiwara-Ohta1996}. Let $p_{k}(z)$ be polynomials defined by
\begin{equation} \label{defpk}
\sum_{k=0}^{\infty}p_k(z) \epsilon^k =\exp\left( z \epsilon - \frac{4}{3}\epsilon^3 \right).
\end{equation}
Then, Yablonskii-Vorob'ev polynomials $Q_{N}(z)$ are given by the $N \times N$ determinant \cite{Kajiwara-Ohta1996}
\begin{eqnarray}  \label{defQN}
&& Q_{N}(z) = c_{N} \left| \begin{array}{cccc}
         p_{1}(z) & p_{0}(z) & \cdots &  p_{2-N}(z) \\
         p_{3}(z) & p_{2}(z) & \cdots &  p_{4-N}(z) \\
        \vdots& \vdots & \vdots & \vdots \\
         p_{2N-1}(z) & p_{2N-2}(z) & \cdots &  p_{N}(z)
       \end{array}
 \right|,
\end{eqnarray}
where $c_{N}= \prod_{j=1}^{N}(2j-1)!!$, and $p_{k}(z)\equiv 0$ if $k<0$. This determinant is a Wronskian since one can see from Eq.~(\ref{defpk}) that $
p'_{k+1}(z)=p_k(z)$, where the prime represents differentiation. Yablonskii-Vorob'ev polynomials are monic polynomials with integer coefficients \cite{Clarkson2003-II}, and the first four of them are
\begin{eqnarray*}
&& Q_1(z)=z, \\
&& Q_2(z)=z^3+4, \\
&& Q_3(z)=z^6 + 20z^3 - 80,  \\
&& Q_4(z)=z(z^9 + 60z^6 + 11200).
\end{eqnarray*}

Root structures of these polynomials have been studied in \cite{Fukutani,Taneda,Clarkson2003-II,Miller2014,Bertola2016}, and the following facts are known.
\begin{enumerate}
\item The degree of the $Q_N(z)$ polynomial is $N(N+1)/2$, which can be easily seen from Eq.~(\ref{defQN}).
\item All roots of $Q_N(z)$ are simple \cite{Fukutani}. Thus, $Q_N(z)$ has $N(N+1)/2$ simple roots.
\item Zero is a root of $Q_N(z)$ if and only if $N\equiv 1 \hspace{0.1cm} \mbox{mod} \hspace{0.1cm} 3$ \cite{Taneda}.
\item $Q_N(z)$ can be factorized as $Q_N(z)=z^m f(\zeta)$, where $m=1$ when $N\equiv 1 \hspace{0.1cm} \mbox{mod} \hspace{0.1cm} 3$ and $m=0$ otherwise, $\zeta \equiv z^3$, and $f(\zeta)$ is a polynomial of $\zeta$ with integer coefficients and a nonzero constant term \cite{Clarkson2003-II}. This factorization shows that the root structure of $Q_N(z)$ is invariant under $120^\circ$-angle rotation in the complex $z$ plane.
\item Roots of $Q_N(z)$ exhibit a triangular pattern in the complex plane for all $N\ge 2$ \cite{Clarkson2003-II,Miller2014,Bertola2016}. This fact is not surprising given the $120^\circ$ rotational symmetry of $Q_N(z)$'s root structure mentioned above.
\item Roots of $Q_N(z)$ are also symmetric with respect to the real-$z$ axis, since the coefficients of $Q_N(z)$ are real and thus complex roots appear in conjugate pairs. This conjugate symmetry, together with the $120^\circ$ rotational symmetry, implies that one vertex of the triangular root structure of $Q_N(z)$ is on the real-$z$ axis.
\end{enumerate}
Due to importance of these root structures to our work, we reproduce some of them in Fig. 1 for $2\le N\le 5$.

\begin{figure}[htb]
\begin{center}
\includegraphics[scale=0.45, bb=550 0 355 305]{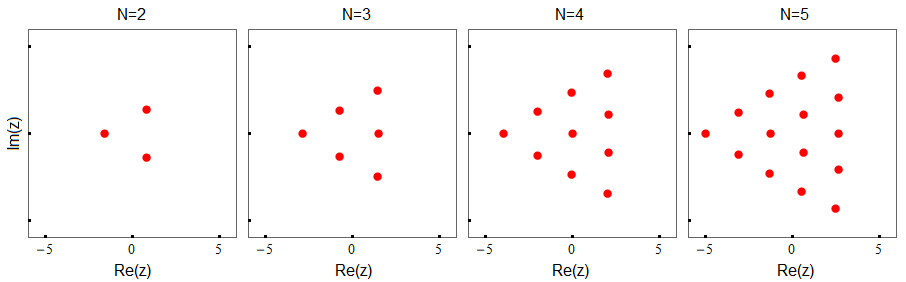}
\caption{Root structures of Yablonskii-Vorob'ev polynomials $Q_{N}(z)$ in the complex $z$ plane for $2\le N\le 5$.}
\end{center}
\end{figure}

\subsubsection{Wronskian-Hermit polynomials and their root structures} \label{secWH}
Next, we introduce Wronskian-Hermit polynomials. Let $q_{k}(z)$  be polynomials defined by
\[\label{qk}
\sum_{k=0}^{\infty} q_{k}(z) \epsilon^k= \exp \left(z \epsilon + \epsilon^2 \right).
\]
These $q_{k}(z)$ polynomials are related to Hermit polynomials through simple variable scalings. Then, for any positive integer $N$ and index vector $\Lambda=(n_1, n_2, \dots, n_N)$, where $\{n_i\}$ are positive and distinct integers in ascending order, i.e., $n_1<n_2<\cdots <n_N$,  the Wronskian-Hermite polynomial $W_{\Lambda}(z)$ is defined as the Wronskian of $q_k(z)$ polynomials
\[\label{WronskianHermite}
W_{\Lambda}(z)=\mbox{Wron}\left[ q_{n_1}(z), q_{n_2}(z),\ldots, q_{n_N}(z) \right],
\]
or equivalently,
\[  \label{defWH}
W_{\Lambda}(z)=\left| \begin{array}{cccc}
         q_{n_1}(z) & q_{n_1-1}(z) & \cdots &  q_{n_1-N+1}(z) \\
         q_{n_2}(z) & q_{n_2-1}(z) & \cdots &  q_{n_2-N+1}(z) \\
        \vdots& \vdots & \vdots & \vdots \\
         q_{n_N}(z) & q_{n_N-1}(z) & \cdots &  q_{n_N-N+1}(z)
       \end{array}
 \right|,
\]
since we can see $q_{k+1}'(z)=q_k(z)$ from the definition (\ref{qk}). In the above determinant, $q_{k}(z)\equiv 0$ when $k<0$.

Regarding root structures of Wronskian-Hermite polynomials $W_{\Lambda}(z)$, we have the following facts.
\begin{enumerate}
\item The degree of the polynomial $W_{\Lambda}(z)$ is equal to $\rho$, where $\rho$ is given in Eq.~(\ref{defrho}). This fact can be seen from the definition (\ref{defWH}).
\item The multiplicity of the zero root in $W_{\Lambda}(z)$ is equal to $d(d+1)/2$, where
\[ \label{defd}
d=k_{odd}-k_{even},
\]
and $k_{odd}$, $k_{even}$ are the numbers of odd and even elements in the index vector $(n_1, n_2, \dots, n_N)$ respectively. This fact was mentioned in~\cite{Felder2012,Garcia_Ferrero2015} and proved in~\cite{Bonneux2020}. If $d(d+1)/2=0$, i.e., $d=0$ or $-1$, then zero is not a root of $W_{\Lambda}(z)$.
\item
The number of nonzero roots (counting multiplicity) in $W_{\Lambda}(z)$, which we denote as $N_{W}$, is
\[ \label{defNW}
N_{W}=\rho-\frac{d(d+1)}{2}.
\]
\item The polynomial $W_{\Lambda}(z)$ can be factored as $W_{\Lambda}(z)=z^{d(d+1)/2} f(\zeta)$, where $d$ is given in Eq.~(\ref{defd}), $\zeta \equiv z^2$, and $f(\zeta)$ is a polynomial of $\zeta$ with real coefficients and a nonzero constant term \cite{Bonneux2020}.
\item If $z_0$ is a root of $W_{\Lambda}(z)$, so are $-z_0, z_0^*$ and $-z_0^*$. This quartet root symmetry can be seen from the above factorization of $W_{\Lambda}(z)$ and the fact that the coefficients of the polynomial $W_{\Lambda}(z)$ are real. As a consequence of this quartet symmetry, the root structure of $W_{\Lambda}(z)$ is non-triangular. This is a big difference from Yablonskii-Vorob'ev polynomials, which feature triangular root structures.
\end{enumerate}
In addition, we have the following lemma.
\begin{quote}
\textbf{Lemma 1.} The Wronskian-Hermite polynomial $W_{\Lambda}(z)$ has only zero roots, i.e., $N_{W}=0$, if and only if  $(n_1, n_2, \dots, n_N)=(1, 3, 5, \cdots, 2N-1)$.
\end{quote}

\noindent
\textbf{Proof.} Since $k_{odd}+k_{even}=N$, we have from Eq.~(\ref{defNW}) that
\[ \label{defNW2}
N_{W} = \sum_{i=1}^N n_i  - k_{odd}^2- k_{even}(k_{even}-1).
\]
Since $\{n_i\}$ are distinct positive integers, their smallest possible values, after reordering, are $\{1, 3, \cdots 2k_{odd}-1, 2, 4, \cdots, 2k_{even}\}$. Thus,
\[
 \sum_{i=1}^N n_i \ge \left[1+3+\cdots (2k_{odd}-1)\right] +\left (2+4+\cdots+2k_{even}\right)=k_{odd}^2+k_{even}(k_{even}+1).
 \]
 Then,
 \[  \label{mW}
 N_{W} \ge 2k_{even}.
 \]
For $W_{\Lambda}(z)$ to have only zero roots, $N_{W}$ must be zero; so $k_{even}=0$, i.e., all numbers in $\{n_i\}$ must be odd. In addition, for the equality in (\ref{mW}) to hold, all these odd and distinct numbers must be the lowest, i.e., $(n_1, n_2, \dots, n_N)=(1, 3, 5, \cdots, 2N-1)$. This completes the proof.

This lemma tells us that, as long as $\Lambda\ne (1, 3, 5, \cdots, 2N-1)$, the Wronskian-Hermit polynomial $W_{\Lambda}(z)$ would always have nonzero roots. This result is important to us, as we will show in later text that the presence or absence of nonzero roots in $W_{\Lambda}(z)$ will have direct consequences on the solution patterns of higher-order lumps.

On roots of Wronskian-Hermite polynomials, beside the above facts, the following conjecture has also been proposed.
\begin{quote}
\textbf{Conjecture 1} \cite{Felder2012}.  All roots of every Wronskian-Hermite polynomial $W_{\Lambda}(z)$ are simple, except possibly the zero root.
\end{quote}
This conjecture will be useful, as we will show in later text that the multiplicity of a root in the Wronskian-Hermite polynomial has direct implications on the wave structure of higher-order lumps. Based on this conjecture, $W_{\Lambda}(z)$ would have $N_W$ nonzero simple roots, where $N_W$ is given in Eq.~(\ref{defNW}). We have checked this conjecture on a number of examples of $W_{\Lambda}(z)$, and found it to always hold.

To illustrate root structures of Wronskian-Hermite polynomials, we choose two index vectors
\[       \label{Lambda12}
\Lambda_1=(2, 3, 4, 5), \quad \Lambda_2=(3, 4, 5, 7, 9).
\]
The corresponding polynomials are found to be
\begin{eqnarray}
&& W_{\Lambda_1}(z)=\frac{z^8-16 z^6+120 z^4+720}{2880}, \label{W1} \\
&& W_{\Lambda_2}(z)=-\frac{z^6 \left(z^{12}-12 z^{10}+180 z^8+672 z^6-7056 z^4-181440 z^2-1270080\right)}{2743372800}.  \label{W2}
\end{eqnarray}
Root structures of these two polynomials are plotted in Fig. 2. It is seen that for the first polynomial, its root structure is rectangular and does not contain zero. For the second polynomial, its root structure is quasi-rectangular with a zero root (of multiplicity six) in the center. All nonzero roots in these two polynomials are simple, which is consistent with the earlier conjecture.

\begin{figure}[htb]
\begin{center}
\includegraphics[scale=0.4, bb=200 0 355 355]{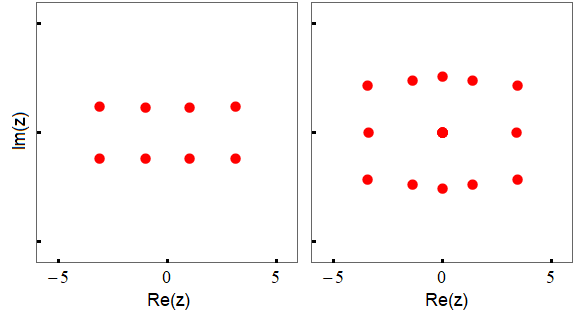}
\caption{Root structures of Wronskian-Hermite polynomials  $W_{\Lambda_1}(z)$ (left) and $W_{\Lambda_2}(z)$ (right) in the complex plane $z$, where
index vectors $\Lambda_1$ and $\Lambda_2$ are given in Eq.~(\ref{Lambda12}). }
\end{center}
\end{figure}

\section{Patterns of higher-order lumps at large times}
In this section, we study patterns of higher-order lumps at large times. In this study, we will set the spectral parameter $p=1$ without loss of generality (see Remark 4 in the previous section). In this case, the constant factor in Eq.~(\ref{Schmatrimnij}) simplifies to $1/4^\nu$. In addition, the definition (\ref{schurcoeffsr}) of the $\textbf{\emph{s}}$ vector reduces to
\begin{eqnarray}
\ln \left[\frac{2}{\kappa}  \tanh \left(\frac{\kappa}{2}\right)\right] = \sum_{k=1}^{\infty}s_{k} \hspace{0.05cm} \kappa^k,  \label{schurcoeffsr2}
\end{eqnarray}
which is identical to the $\textbf{\emph{s}}$ vector in the earlier work \cite{OhtaJY2012} on rogue waves of the NLS equation. In particular, all odd-indexed elements $s_{odd}$ of $\textbf{\emph{s}}$ are zero since the function on the left side of the above equation is even.

It turns out that pattern analysis of lumps depends on whether vectors of internal parameters $\{\textbf{\emph{a}}_{i}\}$ are the same vectors (i.e., whether vector elements $a_{i, j}$ depend on the $i$ index). In this paper, we only consider the case where these $\{\textbf{\emph{a}}_{i}\}$ vectors are the same, i.e., $\textbf{\emph{a}}_{i}=\textbf{\emph{a}}$. In this case, through a shift of the $(x, y)$ axes, we can make the first element of $\textbf{\emph{a}}$ to be zero. Thus, our parameter choices are
\[  \label{cond_a}
\textbf{\emph{a}}_{i}=\textbf{\emph{a}}=(0, a_2, a_3, \cdots).
\]
Under these parameters, we have two theorems on patterns of higher-order lumps at large times, depending on whether the index vector $\Lambda$ is equal to $(1, 3, 5, \dots, 2N-1)$.

\subsection{Large-time lump patterns when  $\Lambda=(1, 3, 5, \dots, 2N-1)$}
Our first theorem is for the case where the index vector $\Lambda$ is equal to $(1, 3, 5, \dots, 2N-1)$.
\begin{quote}
\textbf{Theorem 2.} If the index vector $\Lambda=(1, 3, 5, \dots, 2N-1)$, then, when $|t|\gg 1$,  the higher-order lump solution $u_{\Lambda} (x,y,t)$ asymptotically separates into $N(N+1)/2$ fundamental lumps $u_{1}(x-x_{0}, \hspace{0.04cm}  y-y_{0}, t)$, where $u_1(x,y,t)$ is given in Eq.~(\ref{defu1}),
\[\label{x0t01}
x_0= \Re(z_0) \hspace{0.05cm} (12t)^{1/3}, \quad
y_0=\frac{\Im(z_0)}{2}(12t)^{1/3},
\]
$z_0$ is each of the $N(N+1)/2$ simple roots of the Yablonskii--Vorob'ev polynomial $Q_N(z)$, and $\Re$, $\Im$ represent the real and imaginary parts of a complex number. The peak of each fundamental lump is spatially located at $(x, y)=(12 t+x_0, y_0)$. The absolute error of this fundamental-lump approximation is $O(|t|^{-1/3})$ when $z_0\ne 0$ and $O(|t|^{-1})$ when zero is a root and $z_0=0$. Expressed mathematically, when $(x, y)$ is in the neighborhood of each of these fundamental lumps, i.e., $(x-12 t-x_{0})^2+(y-y_{0})^2=O(1)$, we have the following solution asymptotics for $|t|\gg 1$,
\[ \label{Theorem2asym}
u_{\Lambda}(x,y,t) = \left\{\begin{array}{ll} u_{1}(x-x_{0}, \hspace{0.04cm}  y-y_{0}, t) + O\left(|t|^{-1/3}\right), & \mbox{if}\hspace{0.07cm} z_0\ne 0, \\ u_{1}(x-x_{0}, \hspace{0.04cm}  y-y_{0}, t) + O\left(|t|^{-1}\right), & \mbox{if}\hspace{0.07cm} z_0= 0.  \end{array} \right.
\]
When $(x,y)$ is not in the  neighborhood of any of these $N(N+1)/2$ fundamental lumps, $u_{\Lambda} (x,y,t)$ asymptotically approaches zero as $|t|\to \infty$.
\end{quote}
The proof of this theorem will be provided in Sec.~\ref{sec:proof}.

This theorem indicates that, wave patterns at large times are formed by $N(N+1)/2$  fundamental lumps. Relative to the moving frame of $x$-direction velocity $12 $, positions $(x_0, y_0)$ of these fundamental lumps are just a simple linear transformation of the root structure of the Yablonskii--Vorob'ev polynomial $Q_N(z)$, i.e.,
\[ \label{x0t0B}
\left[\begin{array}{c} x_{0} \\ y_{0} \end{array}\right] = (12t)^{1/3} \left[ \begin{array}{cc} 1 & 0 \\ 0 & \frac{1}{2} \end{array}\right]\left[\begin{array}{c} \Re(z_0) \\ \Im(z_0) \end{array}\right].
\]
Since the transformation matrix is diagonal, this transformation is simply a stretching along both horizontal and vertical directions. Recall that the Yablonskii--Vorob'ev root structure is triangular (see Fig. 1). The resulting lump pattern is then triangular as well. When $t\gg 1$, this triangular lump pattern preserves the same orientation of the original triangle of the Yablonskii--Vorob'ev root structure. But when $t\ll -1$, the triangular lump pattern would be oriented opposite of the Yablonskii--Vorob'ev root structure. Indeed, it is easy to see from Eq.~(\ref{x0t0B}) that, when time changes from large negative to large positive, i.e., from $-t$ to $+t$, their lump positions would be related as
\[
\left[\begin{array}{c} x_{0}^+ \\ y_{0}^+ \end{array}\right] =-\left[\begin{array}{c} x_{0}^- \\ y_{0}^- \end{array}\right].
\]
Thus, these triangular lump patterns have reversed directions along the $x$-axis (the $y$-direction reversal does not matter since the pattern is symmetric in $y$). This $x$-direction reversal of triangular lump patterns when time changes from large negative to large positive is a dramatic pattern transformation in the KP-I equation. This phenomenon has been graphically reported in \cite{Dubard2013} on several low-order solution examples. Here, we established this fact for the general case.

Theorem 2 also indicates that, at large time, fundamental lumps in the solution complex separate from each other in proportion to $|t|^{1/3}$. This rate of separation is very slow, relative to the overall (linear) speed $12 $ of the whole complex.

One more feature of Theorem 2 is that, positions $(x_0, y_0)$ in Eq.~(\ref{x0t01}) for individual fundamental lumps in the solution complex are independent of the solution's internal parameters $\textbf{\emph{a}}$. This means that, when $|t|\to \infty$, solutions $u_{\Lambda} (x,y,t)$ with different internal parameters $\textbf{\emph{a}}$ would approach the same limit solution.

\subsection{Large-time lump patterns when $\Lambda \ne (1, 3, 5, \dots, 2N-1)$}
When $\Lambda \ne (1, 3, 5, \dots, 2N-1)$, the Wronskian-Hermite polynomial $W_{\Lambda}(z)$ has a zero root of multiplicity $d(d+1)/2$, with $d$ given in Eq.~(\ref{defd}), as well as nonzero roots that are conjectured to be all simple (see Sec. \ref{secWH}). Note that the zero root would be absent if $d=0$ or $-1$; but nonzero roots always exist. In this case, our results on solution patterns at large time are summarized in the following theorem.
\begin{quote}
\textbf{Theorem 3. }
Suppose the index vector $\Lambda\ne (1, 3, 5, \dots, 2N-1)$, and all nonzero roots of the Wronskian-Hermite polynomial $W_{\Lambda}(z)$ are simple. Then, for $|t|\gg 1$, the following asymptotics for the solution $u_{\Lambda}(x,y,t)$ holds.
\begin{enumerate}
\item In the outer region --- the region that is $O\left(|t|^{1/2}\right)$ away from the wave center $(x, y)=(12 t, 0)$, or $\sqrt{(x-12 t)^2+y^2}=O\left(|t|^{1/2}\right)$, the higher-order lump $u_{\Lambda} (x,y,t)$ asymptotically separates into $N_W$ fundamental lumps $u_{1}(x-x_{0}, \hspace{0.04cm}  y-y_{0}, t)$, where $N_W$ is given in Eq.~(\ref{defNW}), $u_1(x, y, t)$ is given in Eq.~(\ref{defu1}),
\[\label{x0t02}
x_0= \Re\left[z_0 (-12t)^{1/2}\right] +O(1), \quad
y_0=\frac{\Im\left[z_0 (-12t)^{1/2}\right]}{2} +O(1),
\]
and $z_0$ is each of the $N_W$ nonzero simple roots of $W_{\Lambda}(z)$.
The absolute error of this fundamental-lump approximation is $O(|t|^{-1/2})$. Expressed mathematically,
when $(x, y)$ is in the neighborhood of each of these fundamental lumps, i.e., $(x-12 t-x_{0})^2+(y-y_{0})^2=O(1)$, we have the following solution asymptotics
\[ \label{Theorem3asym}
u_{\Lambda}(x,y,t) =u_{1}(x-x_{0}, \hspace{0.04cm}  y-y_{0}, t) + O\left(|t|^{-1/2}\right), \quad |t|\gg 1.
\]
\item If zero is a root of $W_{\Lambda}(z)$, i.e., $d\ne 0$ and $-1$, then in the inner region --- the region that is within $O(|t|^{1/3})$ of the wave center $(x, y)=(12 t, 0)$, or $\sqrt{(x-12 t)^2+y^2}\le O\left(|t|^{1/3}\right)$, lies $d(d+1)/2$ fundamental lumps $u_{1}(x-x_{0}, \hspace{0.04cm}  y-y_{0}, t)$, where $u_1(x, y, t)$ is given in Eq.~(\ref{defu1}),
\[\label{x0t09}
x_0= \Re(z_0) \hspace{0.05cm} (12t)^{1/3}+O(1), \quad
y_0=\frac{\Im(z_0)}{2}(12t)^{1/3} +O(1),
\]
and $z_0$ is each of the $d(d+1)/2$ simple roots of the Yablonskii--Vorob'ev polynomial $Q_{\hat{d}}(z)$, with $\hat{d}$ defined as
\[ \label{defdhat}
\hat{d}=\left\{ \begin{array}{ll} d, &  \mbox{when} \hspace{0.1cm} d\ge 0, \\  |d|-1, & \mbox{when} \hspace{0.1cm} d\le -1. \end{array}\right.
\]
Notice that $d(d+1)/2=\hat{d}(\hat{d}+1)/2$. The absolute error of this fundamental-lump approximation is $O(|t|^{-1/3})$ when $z_0\ne 0$ and $O(|t|^{-1})$ when zero is a root of $Q_{\hat{d}}(z)$ and $z_0=0$. Expressed mathematically, when $(x, y)$ is in the neighborhood of each of these fundamental lumps, i.e., $(x-12 t-x_{0})^2+(y-y_{0})^2=O(1)$, with $(x_0, y_0)$ given in (\ref{x0t09}), we have the following solution asymptotics for $|t|\gg 1$,
\[ \label{Theorem3asymb}
u_{\Lambda}(x,y,t) = \left\{\begin{array}{ll} u_{1}(x-x_{0}, \hspace{0.04cm}  y-y_{0}, t) + O\left(|t|^{-1/3}\right), &  \mbox{if}\hspace{0.07cm} z_0\ne 0, \\ u_{1}(x-x_{0}, \hspace{0.04cm}  y-y_{0}, t) + O\left(|t|^{-1}\right), &  \mbox{if}\hspace{0.07cm} z_0= 0.  \end{array} \right.
\]
\item When $(x,y)$ is not in the neighborhood of any of the above fundamental lumps, $u_{\Lambda} (x,y,t)$
asymptotically approaches zero as $|t|\to \infty$.
\end{enumerate}
\end{quote}

\textbf{Remark 6.}  In this theorem, we assumed all nonzero roots of $W_{\Lambda}(z)$ simple, which is true for all examples we tested, such as the two in Eqs.~(\ref{W1})-(\ref{W2}). In view of Conjecture 1 in the previous section, this assumption is expected to hold in all cases. If this conjecture is false, i.e., some nonzero roots of $W_{\Lambda}(z)$ are not simple, then this theorem for the outer region, i.e., Eqs.~(\ref{x0t02})-(\ref{Theorem3asym}), would still hold, but only for nonzero simple roots $z_0$ of $W_{\Lambda}(z)$.

Now, we explain what Theorem 3 says regarding solution patterns at large times when $\Lambda \ne (1, 3, 5, \dots, 2N-1)$. In this case, Theorem 3 indicates that, the whole wave field is generically split up into two regions featuring different patterns.

\begin{enumerate}
\item In the outer region --- the region that is $O(|t|^{1/2})$ away from the wave center $(x, y)=(12 t, 0)$, the wave field
at large time comprises $N_W$ fundamental lumps, whose positions are given through the nonzero roots of the Wronskian-Hermite polynomial $W_{\Lambda}(z)$. Specifically, relative to the moving frame of $x$-direction velocity
$12 $, positions $(x_0, y_0)$ of these fundamental lumps, to the leading order of large time, are just a linear transformation of $W_{\Lambda}(z)$'s nonzero-root structure. The reader is reminded from Sec.~\ref{secWH} that when $\Lambda\ne (1, 3, 5, \dots, 2N-1)$, nonzero roots of $W_{\Lambda}(z)$ always exist, and their shape in the $z$-plane is non-triangular.
When $t$ is large negative, these fundamental-lump positions to the leading order are
\[ \label{x0t0C}
\left[\begin{array}{c} x_{0}^{-} \\ y_{0}^{-} \end{array}\right] = (12|t|)^{1/2} \left[ \begin{array}{cc} 1 & 0 \\ 0 & \frac{1}{2} \end{array}\right]\left[\begin{array}{c} \Re(z_0) \\ \Im(z_0) \end{array}\right],
\]
where $z_0$ is any nonzero root of $W_{\Lambda}(z)$. However, when $t$ is large positive, these lump positions become
\[ \label{x0t0D}
\left[\begin{array}{c} x_{0}^{+} \\ y_{0}^{+} \end{array}\right] = (12|t|)^{1/2} \left[ \begin{array}{cc} 0 & -1 \\ \frac{1}{2} &  0 \end{array}\right]\left[\begin{array}{c} \Re(z_0) \\ \Im(z_0) \end{array}\right].
\]
In the former case, the wave pattern formed by these fundamental lumps is simply a stretching of the Wronskian-Hermite nonzero-root structure along both horizontal and vertical directions. But in the latter case, on top of this stretching, the horizontal and vertical directions are also swapped. In both cases, the resulting wave patterns from transformations (\ref{x0t0C})-(\ref{x0t0D}) are non-triangular since the root structure of $W_{\Lambda}(z)$ is non-triangular.

From the above two transformations, we see that fundamental lumps at large negative time $-t$ and large positive time $+t$ in the outer region are related as
\[
\left[\begin{array}{c} x_{0}^+ \\ y_{0}^+ \end{array}\right] =\left[ \begin{array}{cc} 0 & -2 \\ \frac{1}{2} &  0 \end{array}\right]
\left[\begin{array}{c} x_{0}^- \\ y_{0}^- \end{array}\right].
\]
Thus, when time goes from large negative to large positive, outer-region lump patterns in the $(x, y)$ plane have swapped horizontal and vertical directions. In addition, stretching of different amounts has also occurred along the two directions. This swapping of horizontal and vertical directions is another type of dramatic pattern transformation, and it is very different from the triangular $x$-direction reversal that occurs when $\Lambda = (1, 3, 5, \dots, 2N-1)$. For certain single-line patterns of fundamental lumps, a change from a vertical line to a horizontal line in the $(x,y)$ plane has been graphically reported in \cite{Chen2016} and analytically explained in \cite{Chang2018}. Here, we proved this fact for the general case, where patterns of fundamental lumps based on Wronskian-Hermite root structures can be arbitrary, not just lines (see the next section for examples).

In this outer region, fundamental lumps at large time separate from each other in proportion to $|t|^{1/2}$. This is another big difference between the present solutions and those with $\Lambda = (1, 3, 5, \dots, 2N-1)$ in the previous subsection, where lumps separate in proportion to $|t|^{1/3}$ instead.

\item In the inner region --- the region that is within $O(|t|^{1/3})$ of the wave center $(x, y)=(12 t, 0)$, if $d\ne 0$ and
$-1$, then the solution $u_\Lambda(x, y, t)$ at large time would comprise $d(d+1)/2$ fundamental lumps, whose positions are given through roots of the Yablonskii--Vorob'ev polynomial $Q_{\hat{d}}(z)$, with $\hat{d}$ defined in Eq.~(\ref{defdhat}). The reader is reminded that $\hat{d}(\hat{d}+1)/2=d(d+1)/2$. Relative to the moving frame of $x$-direction velocity $12$, positions $(x_0, y_0)$ of these fundamental lumps, to the leading order of large time, are just a linear transformation of $Q_{\hat{d}}(z)$'s root structure, i.e.,
\[ \label{x0t0B2}
\left[\begin{array}{c} x_{0} \\ y_{0} \end{array}\right] = (12t)^{1/3} \left[ \begin{array}{cc} 1 & 0 \\ 0 & \frac{1}{2} \end{array}\right]\left[\begin{array}{c} \Re(z_0) \\ \Im(z_0) \end{array}\right],
\]
where $z_0$ is each of the $d(d+1)/2$ simple roots of $Q_{\hat{d}}(z)$. This lump-position formula in the inner region is very similar to (\ref{x0t0B}) of the $\Lambda= (1, 3, 5, \dots, 2N-1)$ case. Thus, the pattern of these $d(d+1)/2$ fundamental lumps in the inner region at large time is a simple stretching of $Q_{\hat{d}}(z)$'s root structure, and the resulting pattern is triangular if $\hat{d}>1$. In addition, as time evolves from large negative to large positive, these triangular lump patterns would reverse direction along the $x$-axis. Furthermore, fundamental lumps in this inner region separate from each other in proportion to $|t|^{1/3}$ at large time, similar to the $\Lambda = (1, 3, 5, \dots, 2N-1)$ case in Theorem 2. If $\hat{d}=0$, i.e., $d=0$ or $-1$, this inner region would be absent.
\end{enumerate}

The above results reveal that, the pattern of the solution $u_{\Lambda} (x,y,t)$ for $\Lambda\ne (1, 3, 5, \dots, 2N-1)$ at large time is richer, with the outer region exhibiting the non-triangular shape of the stretched nonzero-root structure of the Wronskian-Hermite polynomial $W_{\Lambda}(z)$, and with the inner region exhibiting the triangular shape of the stretched root structure of the Yablonskii--Vorob'ev polynomial $Q_{\hat{d}}(z)$. As time changes from large negative to large positive, the outer pattern swaps horizontal and vertical directions, while the inner pattern reverses the horizontal direction. These different types of pattern transformations in the outer and inner regions of the same solution are fascinating. When $\Lambda\ne (1, 3, 5, \dots, 2N-1)$, the outer pattern is always present since $W_{\Lambda}(z)$ always has nonzero roots, but the inner pattern is present only when $d\ne 0$ and $-1$ and absent otherwise. When $\Lambda= (1, 3, 5, \dots, 2N-1)$, the outer pattern disappears, since $W_{\Lambda}(z)$ has only zero roots (see Lemma 1). In this special case, our results for the inner region in Theorem~3 are consistent with those in Theorem~2. However, Theorem~2 for this special case is stronger, since it shows that positions $(x_0, y_0)$ of fundamental lumps now have no $O(1)$ shifts [see Eq.~(\ref{x0t01})] --- a more accurate prediction than Eq.~(\ref{x0t09}) of Theorem 3 which shows $O(1)$ shifts in general.

In the end, we note that in the earlier work \cite{Ablowitz2000}, it was reported that at large time, fundamental lumps in the higher-order lump complex separate from each other in proportion to $|t|^q$, where $\frac{1}{3}\le q\le \frac{1}{2}$. Our results in Theorems 2 and 3 indicate that this $q$ value can only be $1/3$ or $1/2$, nothing in between.

\section{Comparison between true lump patterns and analytical predictions}
In this section, we compare our analytical predictions of lump patterns with true solutions.

\subsection{Pattern transformation when  $\Lambda =(1, 3, 5, \dots, 2N-1)$}
\begin{figure}[htb]
\begin{center}
\includegraphics[scale=0.35, bb=0 0 1150 490]{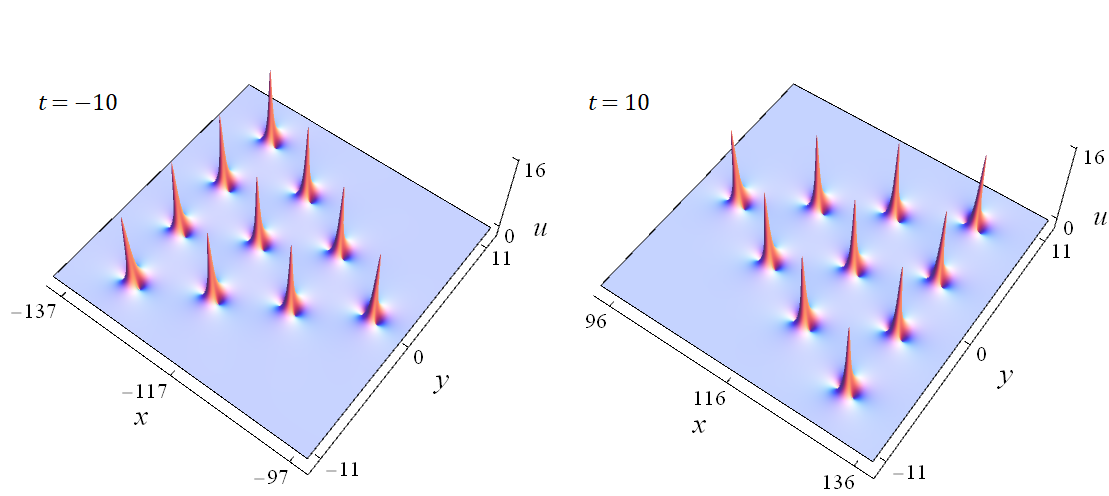}
\caption{Predicted solutions $u_{\Lambda} (x,y,t)$ with $\Lambda =(1, 3, 5, 7)$ at time values of $t=-10$ (left) and $t=10$ (right).}  \label{f:YBp}
\end{center}
\end{figure}

First, we do the comparison when $\Lambda =(1, 3, 5, \dots, 2N-1)$, where a triangular pattern of lumps at large time is predicted. To be specific, we take $N=4$; so $\Lambda =(1, 3, 5, 7)$.
Root structure of the corresponding Yablonskii-Vorob'ev polynomial $Q_{4}(z)$ has been displayed in Fig. 1. Using those roots and formulae (\ref{x0t01}), predicted solutions from Theorem 2 at large times $t=-10$ and $10$ are plotted in Fig. \ref{f:YBp}.

\begin{figure}[htbp]
\begin{center}
\includegraphics[scale=0.35, bb=0 0 1150 467]{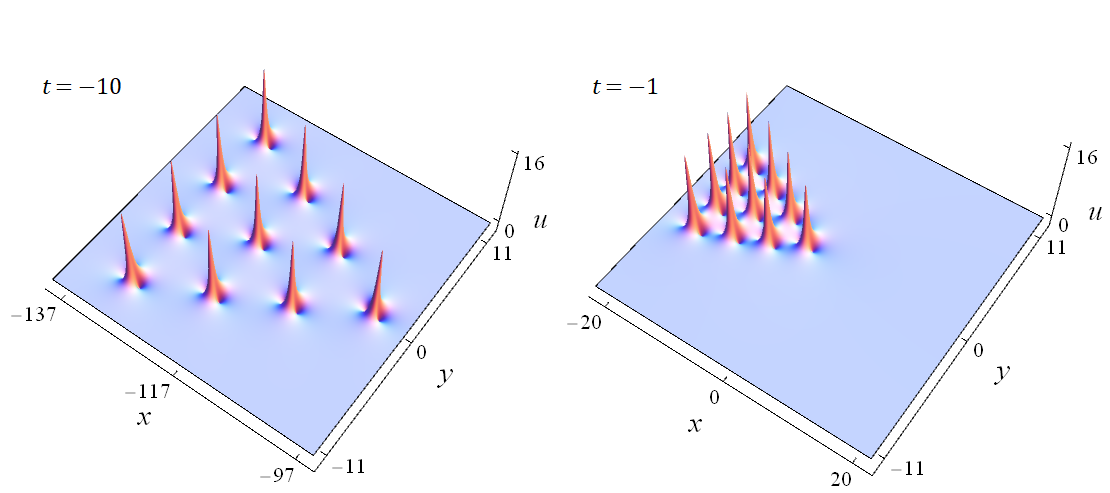}
\includegraphics[scale=0.35, bb=0 0 1150 497]{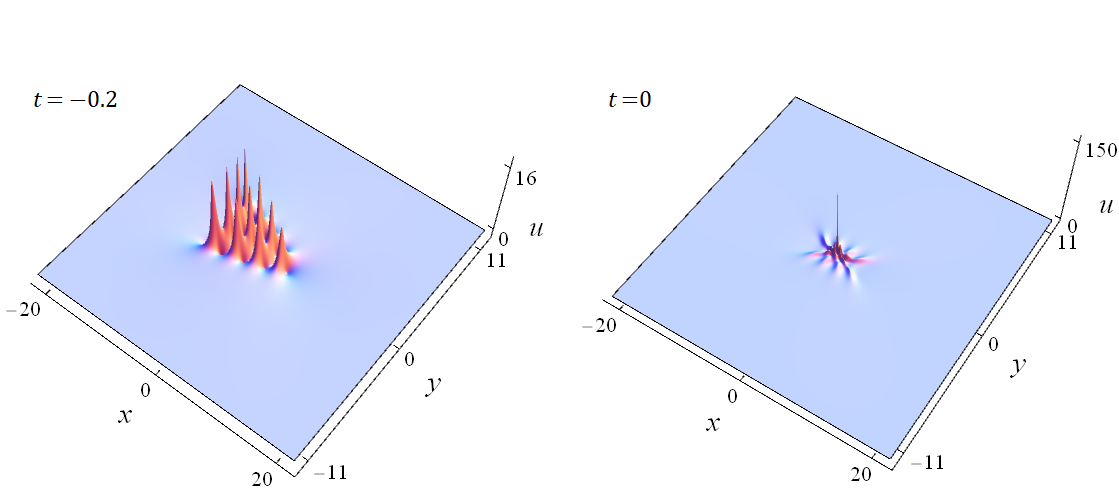}
\includegraphics[scale=0.35, bb=0 0 1150 497]{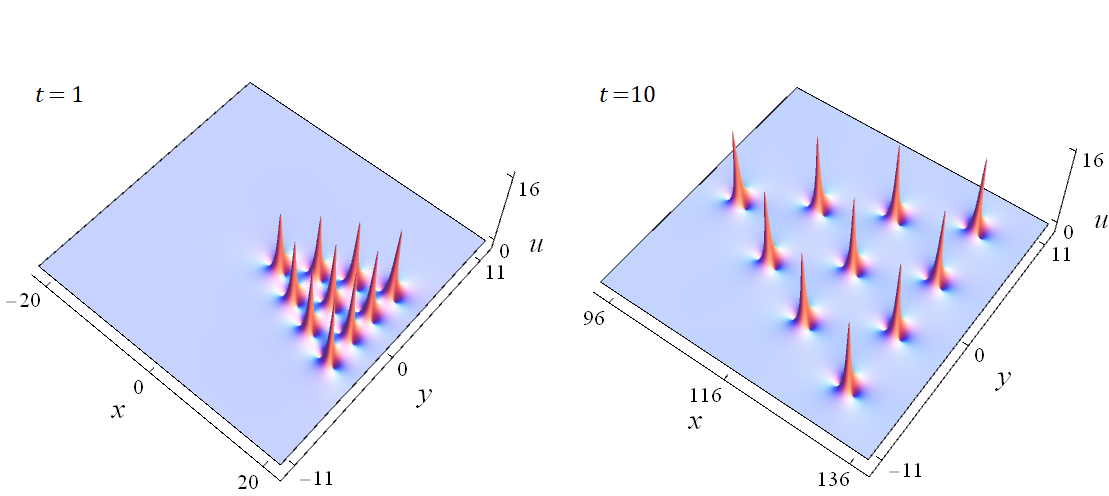}
\caption{The true solution $u_{\Lambda} (x,y,t)$ with $\Lambda =(1, 3, 5, 7)$ and $\textbf{\emph{a}}=(0,0,0,0,0,0,0)$ at various time values shown inside the panels.} \label{f:YBt}
 \end{center}
\end{figure}

Now, we compare these predicted solutions with true ones. In the true solution $u_{\Lambda} (x,y,t)$, we select its internal parameters as $\textbf{\emph{a}}=(0,0,0,0,0,0,0)$. Then, evolutions of this true solution, at six time values of $t=-10,\ -1,\ 0,\ 0.2,\ 1$ and $10$, are plotted in Fig. \ref{f:YBt}. When comparing these true solutions at large times $t=\pm 10$ to those predicted in Fig.~\ref{f:YBp}, they clearly match each other. First, the true solutions at $t=\pm 10$ indeed exhibit a triangular pattern, as the prediction says. Second, the triangular pattern at $t=10$ is indeed a $x$-direction reversal of the triangular pattern at $t=-10$, relative to a frame moving with $x$-direction velocity $12$. Thirdly, we have quantitatively compared the predicted and true solutions at $t=\pm 10$, and found them to match each other as well. For example, we have quantitatively compared the difference between the true peak location of individual lumps and its analytical prediction (\ref{x0t01}), similar to what we did in Figs.~5-6 of Ref.~\cite{YangYang21a} on the error analysis of rogue-pattern predictions. This comparison shows that this difference is indeed $O(|t|^{-1/3})$ when $z_0\ne 0$ and $O(|t|^{-1})$ when $z_0=0$, as our analytical formula (\ref{Theorem2asym}) says. Thus, our asymptotic theory on patterns of higher-order lumps at large times for $\Lambda =(1, 3, 5, \dots, 2N-1)$ is fully confirmed.

In addition to large times, Fig.~\ref{f:YBt} also displays the true solution $u_{\Lambda} (x,y,t)$ at intermediate times, where our asymptotic theory does not apply. These intermediate panels shed light on how the dramatic $x$-direction reversal of triangular patterns takes place as time changes from large negative to large positive. We see that in this solution, as time increases from $-10$ to 10, the triangle of fundamental lumps first approach each other and shrink in size, then coalesce at $t=0$ and form a single lump of extreme height that is ten times that of original fundamental lumps, and then separate into a triangle of fundamental lumps again but with reversed $x$-direction. This transformation process is fascinating.

How will this $u_{\Lambda} (x,y,t)$ solution evolve if its internal parameters $\textbf{\emph{a}}$ are different from $(0,0,0,0,0,0,0)$? Theorem 2 tells us that in this case, the $u_{\Lambda} (x,y,t)$ solution would approach the same asymptotic state as that shown in Fig.~\ref{f:YBt} at large times. At intermediate times, however, this $u_{\Lambda} (x,y,t)$ solution could look very different from that in Fig.~\ref{f:YBt}. For instance, by suitably choosing the $\textbf{\emph{a}}$ values, we can get $u_{\Lambda} (x,y,t)$ solutions whose graphs at $t=0$ exhibit very different patterns such as a pentagon or a heptagon --- a phenomenon that has been reported in \cite{Gaillard2018}. Thus, although these $u_{\Lambda} (x,y,t)$ solutions with different $\textbf{\emph{a}}$ values exhibit the same large-time triangular patterns, how this triangular pattern at large negative time transforms to its $x$-reversed pattern at large positive time is a process that strongly depends on the choices of the internal $\textbf{\emph{a}}$ values.

\subsection{Pattern transformation when  $\Lambda \ne (1, 3, 5, \dots, 2N-1)$}

Next, we perform the comparison when $\Lambda \ne (1, 3, 5, \dots, 2N-1)$, where the solution pattern at large time is determined by nonzero-root structure of the Wronskian-Hermit polynomial $W_\Lambda(z)$ in the outer region, and by root structure of the Yablonskii-Vorob'ev polynomial $Q_{\hat{d}}(z)$ in the inner region (if $\hat{d}>0$). Since this inner region can be present or absent depending on the $d$ value [see Eq.~(\ref{defd})], we will present two examples, one for each case.

\begin{figure}[htb]
\begin{center}
\includegraphics[scale=0.35, bb=550 0 555 425]{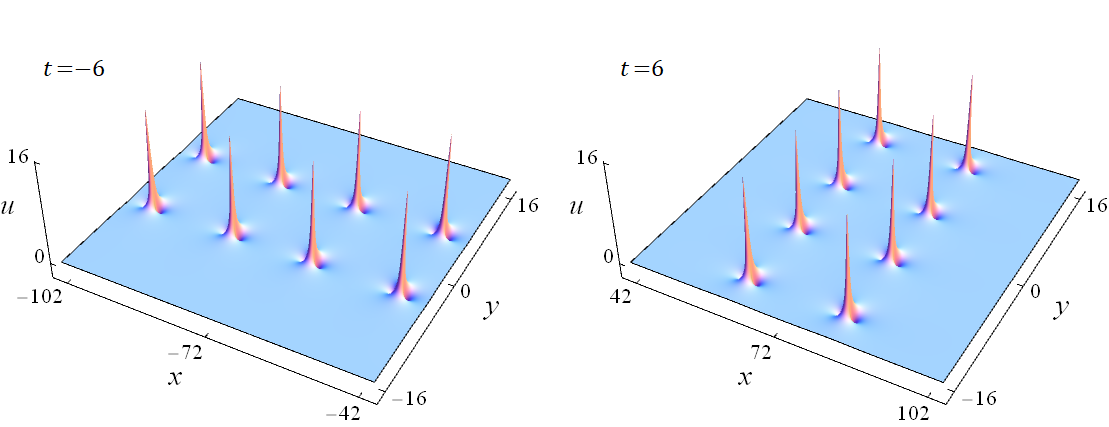}
\caption{Predicted solutions $u_{\Lambda} (x,y,t)$ with $\Lambda =(2, 3, 4, 5)$ at time values of $t=-6$ (left) and $t=6$ (right).}  \label{f:WHp}
\end{center}
\end{figure}

Our first example is $N=4$ and $\Lambda =(2, 3, 4, 5)$. In this case, $d=0$, and thus zero is not a root of $W_\Lambda(z)$ and the inner region is absent. Root structure of the corresponding Wronskian-Hermit polynomial has been displayed in Fig. 2 (the left panel). It was seen that this $W_\Lambda(z)$ admits eight simple nonzero roots which form a rectangle pattern. Using those roots and leading-order terms in formulae (\ref{x0t02}), predicted solutions from Theorem 3 at large times $t=-6$ and $6$ are plotted in Fig. \ref{f:WHp}. The predicted patterns contain eight fundamental lumps which also form a rectangular shape in the $(x, y)$ plane. At $t=-6$, this lump pattern is just a stretching of the Wronskian-Hermit root structure. But at $t=6$, this lump pattern has swapped its $x$ and $y$ directions and changed from its original $x$-direction orientation to the new $y$-direction orientation.

\begin{figure}[htbp]
\begin{center}
\includegraphics[scale=0.35, bb=0 0 1150 430]{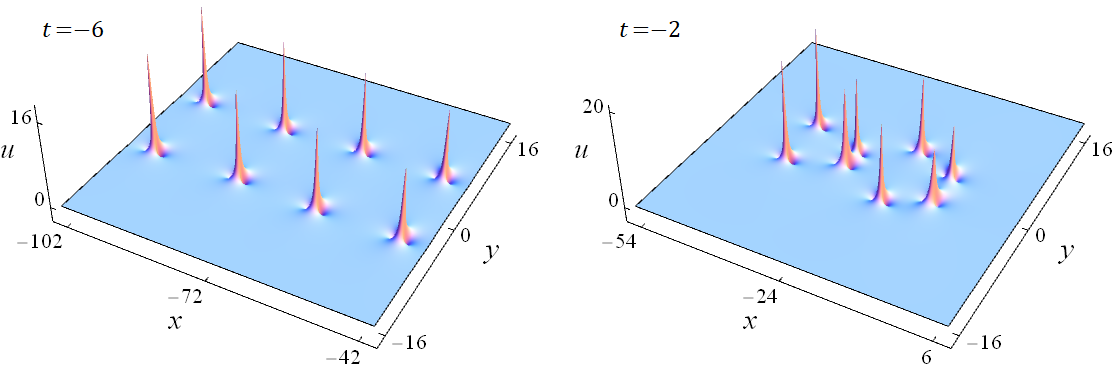}
\includegraphics[scale=0.35, bb=0 0 1150 430]{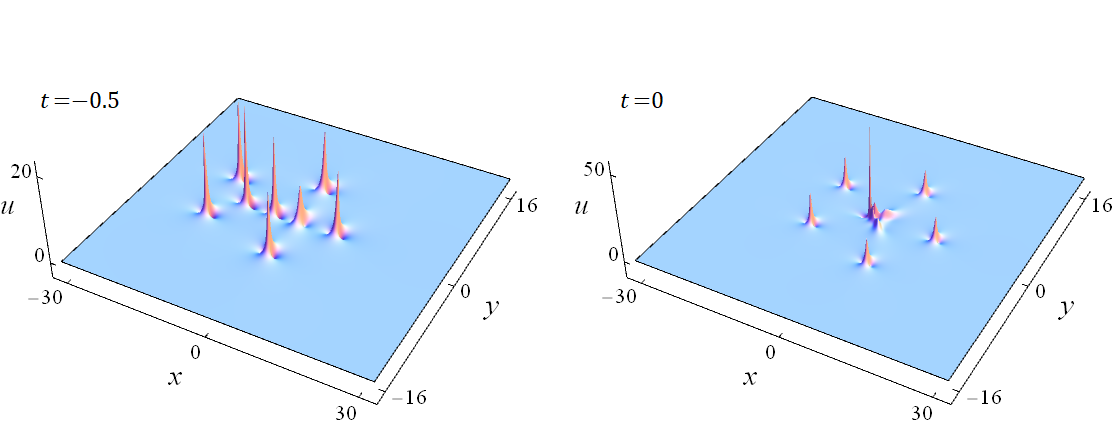}
\includegraphics[scale=0.35, bb=0 0 1150 427]{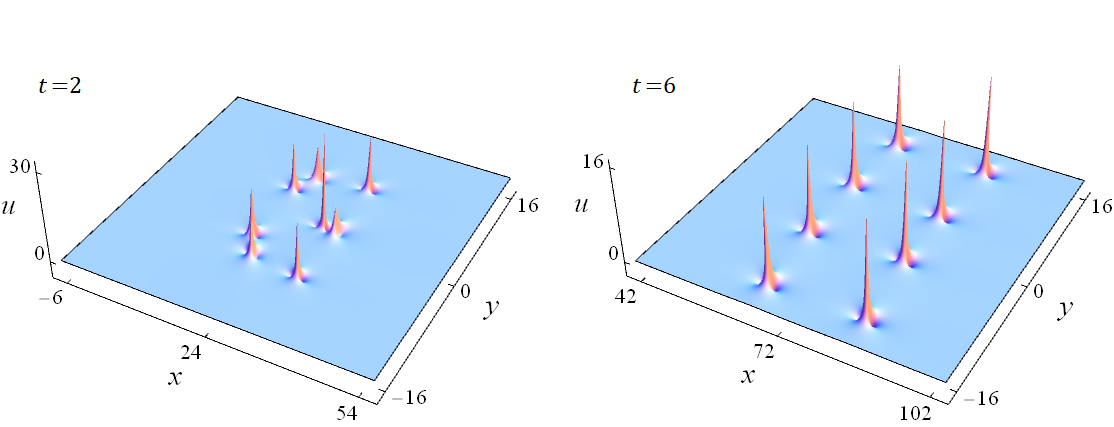}
\caption{True solutions $u_{\Lambda} (x,y,t)$ with $\Lambda =(2, 3, 4, 5)$ and $\textbf{\emph{a}}=(0, 0, 0, 0, 800)$, at various times whose values are shown inside the panels. } \label{f:WHt}
 \end{center}
\end{figure}

To confirm these asymptotic predictions, we plot in Fig.~\ref{f:WHt} the corresponding true solution $u_{\Lambda} (x,y,t)$ at six time values of $t= -6, -2, -0.5, 0, 2$ and $6$. In this true solution, we have selected its internal parameters as $\textbf{\emph{a}}=(0, 0, 0, 0, 800)$. It is seen that at large times of $t=\pm 6$, the true solutions indeed comprise eight fundamental lumps forming a rectangular shape, and their orientations have changed from the $x$-direction to the $y$-direction,
exactly as our asymptotic theory has predicted. In addition, quantitative comparisons between these true rectangular patterns and the predicted ones in Fig.~\ref{f:WHp} show good agreement. Thus, our asymptotic theory on patterns of higher-order lumps at large times is fully confirmed for $\Lambda =(2, 3, 4, 5)$.

By inspecting Fig.~\ref{f:WHt}, we can also see how this dramatic rectangular-pattern reorientation takes place as time increases. First, these eight fundamental lumps of rectangular shape with $x$-direction orientation get closer to each other and rearrange their shapes. At $t=0$, the solution has evolved into a pentagon of five fundamental lumps surrounding a higher-peak lump near the center. Afterwards, this pentagon structure further adjusts its shape in significant ways, until eight new fundamental lumps emerge as a rectangular with $y$-direction orientation in the end. Again, this transformation process is amazing.

\begin{figure}[htb]
\begin{center}
\includegraphics[scale=0.35, bb=0 0 1150 490]{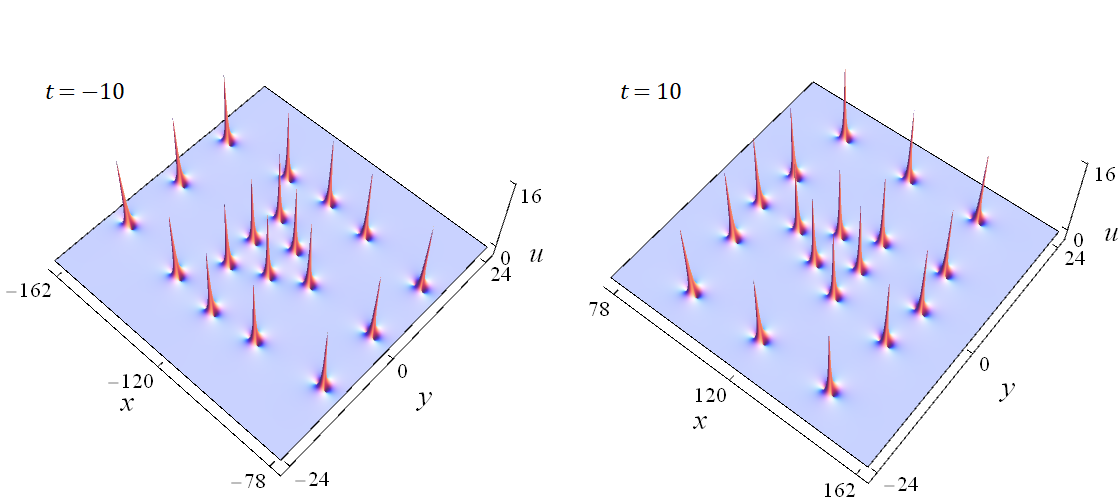}
\caption{Predicted solutions $u_{\Lambda} (x,y,t)$ with $\Lambda =(3, 4, 5, 7, 9)$ at time values of $t=-10$ (left) and $t=10$ (right).} \label{f:WHp2}
\end{center}
\end{figure}

Our second example is $N=5$ and $\Lambda =(3, 4, 5, 7, 9)$. In this case, $d=3$, and thus zero is a root of multiplicity six in $W_\Lambda(z)$, and the inner region is present. Root structure of the corresponding Wronskian-Hermit polynomial has been displayed in Fig. 2 (the right panel). It is seen that this $W_\Lambda(z)$ admits 12 simple nonzero roots which form a quasi-rectangular shape, plus the zero root of multiplicity six at the center of the quasi-rectangle. Using those roots and leading-order terms in formulae (\ref{x0t02}) and (\ref{x0t09}), predicted solutions from Theorem 3 at large times $t=-10$ and $10$ are plotted in Fig.~\ref{f:WHp2}. The predicted patterns contain 12 fundamental lumps which also form a quasi-rectangular pattern in the outer region of the $(x, y)$ plane, plus six fundamental lumps which form a triangle in the inner region. At $t=-10$, the outer lump pattern is a stretching of the Wronskian-Hermit polynomial $W_\Lambda(z)$'s nonzero-root structure, while the inner lump pattern is a stretching of the Yablonskii--Vorob'ev polynomial $Q_{3}(z)$'s root structure. At $t=10$, however, the predicted outer lump pattern has rotated by $90^\circ$ from its $t=-10$ state [plus additional $(x,y)$-direction stretching], while the predicted inner triangular lump pattern has reversed its direction along the $x$-axis.

\begin{figure}[htbp]
\begin{center}
\includegraphics[scale=0.35, bb=0 0 1150 430]{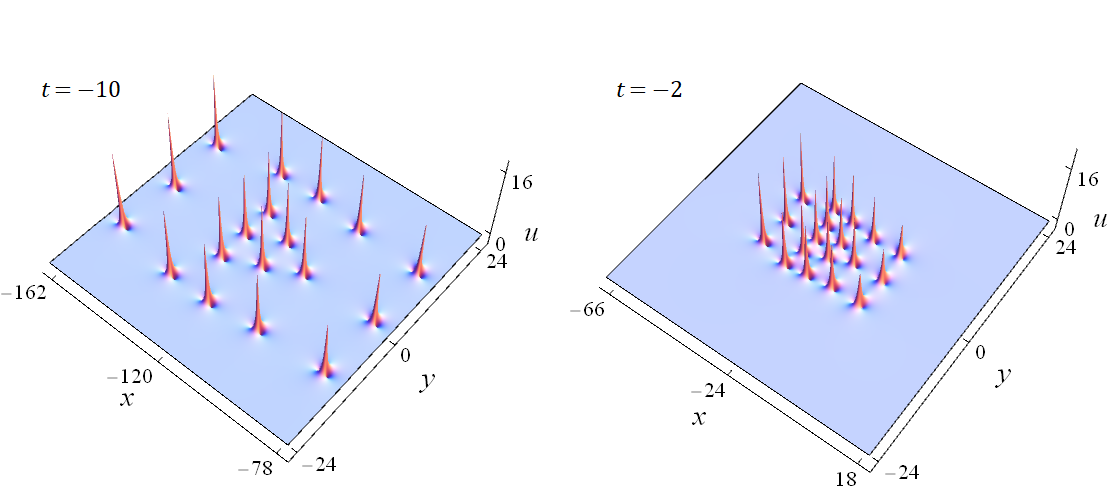}
\includegraphics[scale=0.35, bb=0 0 1150 490]{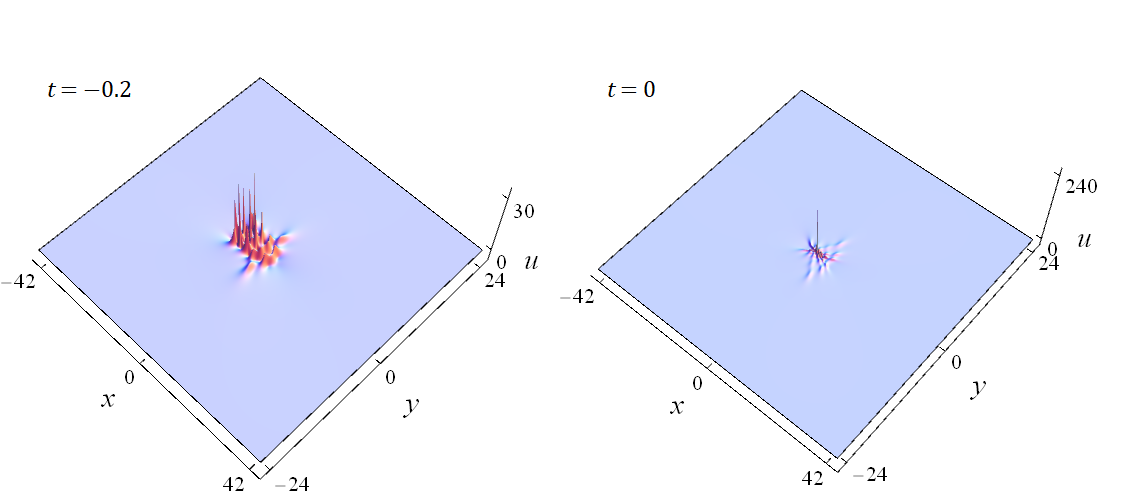}
\includegraphics[scale=0.35, bb=0 0 1150 495]{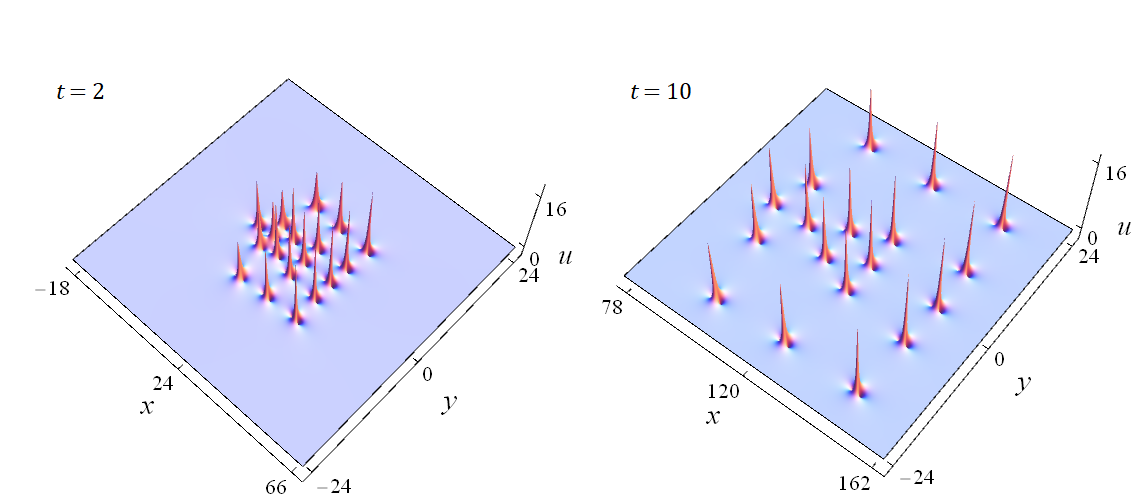}
\caption{True solutions $u_{\Lambda} (x,y,t)$ with $\Lambda =(3, 4, 5, 7, 9)$ and $\textbf{\emph{a}}=(0,0,0,0,0,0,0,0,0)$, at various times whose values are shown inside the panels. }
\label{f:WHt2}
\end{center}
\end{figure}

To confirm these asymptotic predictions, we plot in Fig.~\ref{f:WHt2} the corresponding true solution $u_{\Lambda} (x,y,t)$ at six time values of $t= -10, -2, -0.2, 0, 2$ and $10$. In this true solution, we have selected all-zero internal parameters of $\textbf{\emph{a}}=(0, 0, 0, 0, 0, 0, 0, 0, 0)$. It is seen that at large times of $t=\pm 10$, the true solutions closely match our predictions in the previous figure. Specifically, the true solutions at these large times also split into outer and inner regions, with outer quasi-rectangular patterns and inner triangular patterns closely resembling our predicted ones in Fig.~\ref{f:WHp2}. Quantitative comparisons between these true patterns and predicted ones show good agreement as well. Thus, our asymptotic theory on higher-order lump patterns at large times is fully confirmed for $\Lambda =(3, 4, 5, 7, 9)$.

True solution graphs at intermediate time values in Fig.~\ref{f:WHt2} reveal how these striking pattern transformations in outer and inner regions take place. It is seen that all fundamental lumps in the inner and outer regions at large negative time first move toward each other. Then they merge and coalesce at $t\approx 0$. Afterwards, all these fundamental lumps re-emerge and move away from each other, but not returning to their pre-merging state. Instead, the quasi-rectangular outer lumps have swapped their $x$ and $y$ directions, and the triangular inner lumps have reversed the $x$-direction. These pattern transformations are visually miraculous and mysterious. But due to our Theorem 3, they can now be completely understood from a mathematical point of view.

\section{Proofs of the two theorems} \label{sec:proof}
Now, we prove our two theorems stated in Sec.~3. The reader is reminded that in these proofs, $p=1$ and $\textbf{\emph{a}}_{i}$ are chosen as (\ref{cond_a}) in the higher-order lump solutions of Theorem 1, for reasons which have been explained earlier in the paper. Thus, solution expressions in Theorem 1 can be simplified. Notably, the constant factor in Eq.~(\ref{Schmatrimnij}) simplifies to $1/4^\nu$, and the $\textbf{\emph{s}}$ vector is real with $s_{odd}=0$ (see the beginning of Sec.~3).

\subsection{Proof of Theorem 2}
In this case, $\Lambda=(1, 3, 5, \dots, 2N-1)$. First, we rewrite the determinant (\ref{Blockmatrix}) as a larger $3N\times 3N$ determinant \cite{OhtaJY2012,YangYang21a}
\[ \label{3Nby3Ndet}
\sigma=\left|\begin{array}{cc}
\textbf{O}_{N\times N} & \Phi_{N\times 2N} \\
-\Psi_{2N\times N} & \textbf{I}_{2N\times 2N} \end{array}\right|,
\]
where
\[
\Phi_{i,j}=2^{-(j-1)} S_{2i-j}\left[\textbf{\emph{x}}^{+} + (j-1) \textbf{\emph{s}} +\textbf{\emph{a}}\right], \quad \Psi_{i,j}=2^{-(i-1)} S_{2j-i}\left[(\textbf{\emph{x}}^{+})^* + (i-1) \textbf{\emph{s}}+\textbf{\emph{a}}^*\right],
\]
$S_{j}\equiv 0$ for $j<0$, and vectors $\textbf{\emph{x}}^{+}$ and $\textbf{\emph{s}}$ are given in Eqs.~(\ref{defxrp}) and (\ref{schurcoeffsr2}). This determinant can be further simplified. Indeed, using the technique outlined in Appendix A of Ref.~\cite{YangYang21a}, we can eliminate all $x_{even}^+$ and $a_{even}$ terms from the vectors $\textbf{\emph{x}}^{+}$ and $\textbf{\emph{a}}$, and reduce the above matrix element formulae to
\[ \label{PhiPsi0}
\Phi_{i,j}=2^{-(j-1)} S_{2i-j}\left[\hat{\textbf{\emph{x}}}^{+} + (j-1) \textbf{\emph{s}} +\hat{\textbf{\emph{a}}}\right], \quad \Psi_{i,j}=2^{-(i-1)} S_{2j-i}\left[(\hat{\textbf{\emph{x}}}^{+})^* + (i-1) \textbf{\emph{s}}+\hat{\textbf{\emph{a}}}^*\right],
\]
where
\[ \label{defxahat}
\hat{\textbf{\emph{x}}}^{+} \equiv \left(x_1^+, 0, x_3^+, 0, x_5^+, 0, \cdots\right), \quad
\hat{\textbf{\emph{a}}} \equiv \left(0, 0, a_3, 0, a_5, 0, \cdots\right).
\]
The elimination of the solution's dependence on $x_2^+$ is a key feature of the index vector $\Lambda=(1, 3, 5, \dots, 2N-1)$, and this feature is responsible for the distinctive pattern behaviors described in Theorem 2.

Now, we analyze the large-time asymptotics of the above determinant $\sigma$. For this purpose, we introduce a moving $x$-frame coordinate
\[ \label{defxhat}
\hat{x}\equiv x-12 t.
\]
Then, the elements $x_{k}^+$ in Eq.~(\ref{defxrp}) become
\[ \label{defx1x3}
x_1^+=\hat{x}+2\textrm{i}  y,    \quad  x_k^+=\frac{1}{k!}\hat{x}+\frac{2^k}{k!} \textrm{i}y+T_k,
\]
where $T_k\equiv 12(1-3^{k-1})t/k!$. In particular,
\[ \label{T2T3}
T_2=-12t, \quad T_3=-16t.
\]
In this moving $x$-frame, when $|t|$ is large and $\sqrt{\hat{x}^2+y^2}=O(|t|^{1/3})$, we have the leading-order asymptotics for $S_{k}\left(\hat{\textbf{\emph{x}}}^{+} + \nu \textbf{\emph{s}} +\hat{\textbf{\emph{a}}}\right)$ as
\[ \label{Skasym1}
S_{k}\left(\hat{\textbf{\emph{x}}}^{+} + \nu  \textbf{\emph{s}} +\hat{\textbf{\emph{a}}}\right) \sim S_k(\textbf{v}), \quad \quad |t|  \gg 1,
\]
where
\[  \label{vdef}
\textbf{v}=\left(x_1^+, \hspace{0.04cm} 0,  \hspace{0.04cm} T_3, \hspace{0.04cm} 0,  \hspace{0.04cm} 0, \hspace{0.04cm} 0, \cdots\right).
\]
By comparing the definition of Schur polynomials $S_k(\textbf{v})$ to the definition of $p_k(z)$ polynomials in Eq.~(\ref{defpk}), we see that
\begin{equation} \label{Skorder}
S_k(\textbf{v})=\left(-3T_3/4\right)^{k/3}\hspace{-0.1cm} p_{k}(z),
\end{equation}
where
\[ \label{defz1}
z=(-3T_3/4)^{-1/3}x_1^+=(-3T_3/4)^{-1/3}\left(\hat{x}+2\textrm{i}  y\right).
\]
Using these formulae and the Laplace expansion of the $3N\times 3N$ determinant (\ref{3Nby3Ndet})
\begin{eqnarray} \label{sigmanLap}
&& \hspace{-0.8cm} \sigma=\sum_{0\leq\nu_{1} < \nu_{2} < \cdots < \nu_{N}\leq 2N-1}
\det_{1 \leq i, j\leq N} \left[\frac{1}{2^{\nu_j}} S_{2i-1-\nu_j}(\hat{\textbf{\emph{x}}}^{+}+\nu_j \textbf{\emph{s}}+\hat{\textbf{\emph{a}}}) \right]  \times \det_{1 \leq i, j\leq N}\left[\frac{1}{2^{\nu_j}}S_{2i-1-\nu_j} [(\hat{\textbf{\emph{x}}}^{+})^*+ \nu_j \textbf{\emph{s}}+\hat{\textbf{\emph{a}}}^* ]\right],
\end{eqnarray}
together with the fact that the highest order term of $|t|$ in this $\sigma$ comes from the index choices of $\nu_{j}=j-1$, we can readily show that the highest $t$-power term of $\sigma$ is
\begin{equation} \label{sigmanmax}
\sigma \sim |\alpha_0|^2 \hspace{0.05cm} \left|3T_3/4\right|^{\frac{N(N+1)}{3}} \left|Q_{N}(z)\right|^2,
\quad \quad |t| \gg 1,
\end{equation}
where $\alpha_0=2^{-N(N-1)/2}c_N^{-1}$. Inserting this leading-order term of $\sigma$ into Eq.~(\ref{Schpolysolu}), we see that the solution $u_{\Lambda} (x,y,t)$ approaches zero when $|t|\to \infty$, except at or near $(\hat{x}, y)$ locations $\left(x_0, y_{0}\right)$, i.e., at or near $(x, y)$ locations $\left(12 t+x_0, y_{0}\right)$, where
\[  \label{z0def}
z_0=(-3T_3/4)^{-1/3}\left(x_0 +2\textrm{i}  y_0\right)
\]
is a root of the polynomial $Q_{N}(z)$. Solving this equation, we get the $(x_0, y_0)$ locations given by Eq.~(\ref{x0t01}) in Theorem 2. Due to our requirement of $\sqrt{\hat{x}^2+y^2}=O(|t|^{1/3})$, $z_0$ in the above equation should be nonzero.

In order to derive the solution behavior near this $(x, y)=\left(12 t+x_0, y_{0}\right)$ location, we need to perform a more refined asymptotic analysis and calculate the next-order terms in $t$, since the leading-order term in Eq.~(\ref{sigmanmax}) vanishes at this point. Recalling $s_1=0$, this refined analysis is very similar to that we did for rogue waves in the NLS equation \cite{YangYang21a}. For $z_0\ne 0$ in the $(x_0, y_0)$ formula (\ref{x0t01}), i.e., if the $\left(12 t+x_0, y_{0}\right)$ location is $O(|t|^{1/3})$ away from the wave center $(12 t, 0)$, then in the $O(1)$ neighborhood of $\left(12 t+x_0, y_{0}\right)$, i.e., when $(x-12 t-x_{0})^2+(y-y_{0})^2=O(1)$, we have an asymptotics more refined than (\ref{Skasym1}), which is
\[ \label{Skasym2}
S_{k}\left(\hat{\textbf{\emph{x}}}^{+} + \nu  \textbf{\emph{s}} +\hat{\textbf{\emph{a}}}\right) = S_k(\textbf{v})\left[
1+O(|t|^{-2/3})\right].
\]
This $O(|t|^{-2/3})$ relative error is due to our omission of $\hat{x}/6+4\textrm{i} y/3$ relative to $T_3$ in $x_3^+$, and omission of $x_5^+$ relative to $x_3^+$. Using this refined asymptotics and repeating the same steps as in \cite{YangYang21a}, we find that
\begin{eqnarray}  \label{sigmaTh2asym1}
&& \sigma(x,y,t) = \left|\alpha_0\right|^2 \hspace{0.06cm} \left|Q_{N}'(z_0)\right|^2 |3T_3/4|^{\frac{N(N+1)-2}{3}}\left[ \left(x-12  t -x_0\right)^2+4  (y-y_0)^2+ \frac{1}{4}\right]
\left[1+O\left(|t|^{-1/3}\right)\right],
\end{eqnarray}
where $\alpha_0$ is given below Eq.~(\ref{sigmanmax}). For Yablonskii-Vorob'ev polynomials $Q_{N}(z)$, all roots are simple. Thus, $Q_{N}'(z_0)\ne 0$.

In the $O(1)$ neighborhood of the wave center $\left(12 t, 0\right)$, where $(x-12 t)^2+y^2=O(1)$, we need to perform a separate asymptotic analysis, because the earlier $S_k$ asymptotics (\ref{Skasym1}) and (\ref{Skasym2}) do not hold in this region. In this case, due to Eq.~(\ref{defx1x3}), when we lump $T_{2k+1}$ and $a_{2k+1}$ together in Eq.~(\ref{PhiPsi0}) and recall $T_{2k+1}$ is proportional to $t$, the large-time analysis of the present $\sigma$ determinant (\ref{3Nby3Ndet}) is almost identical to that in Appendix C of Ref.~\cite{YangYang21a} for the analysis of NLS rogue patterns when its internal parameters $(a_3, a_5, \cdots)$ are all large and of the same order. Repeating that analysis, we find that if zero is a root of the Yablonskii-Vorob'ev polynomial $Q_{N}(z)$, i.e., $N\equiv 1 \hspace{0.1cm} \mbox{mod} \hspace{0.1cm} 3$, then
\begin{eqnarray} \label{sigmaTh2asym2}
&& \sigma(x,y,t) = \beta_0 \hspace{0.05cm} |t|^{\frac{N(N+1)-2}{3}}\left[ \left(x-12  t\right)^2+4  y^2+ \frac{1}{4}\right]
\left[1+O\left(|t|^{-1}\right)\right],
\end{eqnarray}
where $\beta_0$ is a $N$-dependent positive constant. If zero is not a root of $Q_{N}(z)$, then
$\sigma(x,y,t)\sim \beta_0 \hspace{0.05cm} |t|^{\frac{(N+2)(N-1)}{3}}$.

Substituting the above two $\sigma$ asymptotics (\ref{sigmaTh2asym1})-(\ref{sigmaTh2asym2}) into the solution expression (\ref{Schpolysolu}) and performing a little simplification, we then get the asymptotics (\ref{Theorem2asym}). Theorem 2 is then proved.

\subsection{Proof of Theorem 3}
In this case, $\Lambda\ne (1, 3, 5, \dots, 2N-1)$. We first rewrite the determinant $\sigma$ in (\ref{Blockmatrix}) as a larger $(N+n_N+1)\times (N+n_N+1)$ determinant
\[ \label{3Nby3Ndet2}
\sigma=\left|\begin{array}{cc}
\textbf{O}_{N\times N} & \Phi_{N\times (n_N+1)} \\
-\Psi_{(n_N+1)\times N} & \textbf{I}_{(n_N+1)\times (n_N+1)} \end{array}\right|,
\]
where
\[
\Phi_{i,j}=2^{-(j-1)} S_{n_i+1-j}\left[\textbf{\emph{x}}^{+} + (j-1) \textbf{\emph{s}} +\textbf{\emph{a}}\right], \quad \Psi_{i,j}=2^{-(i-1)} S_{n_j+1-i}\left[(\textbf{\emph{x}}^{+})^* + (i-1) \textbf{\emph{s}}+\textbf{\emph{a}}^*\right],
\]
and vectors $\textbf{\emph{x}}^{+}$ and $\textbf{\emph{s}}$ are given in Eqs.~(\ref{defxrp}) and (\ref{schurcoeffsr2}). Unlike the previous case, we cannot eliminate $x_2^+$ from this solution now. Our large-time asymptotics of this determinant proceeds as follows.

\subsubsection{Proof for the outer region}
First, we prove the asymptotics (\ref{x0t02})-(\ref{Theorem3asym}) for the outer region. In this region, $\sqrt{\hat{x}^2+y^2}=O(|t|^{1/2})$. Thus, we have the leading-order asymptotics for $S_{k}\left(\textbf{\emph{x}}^{+} + \nu \textbf{\emph{s}} +\textbf{\emph{a}}\right)$ as
\[ \label{Skxnasym1}
S_{k}\left(\textbf{\emph{x}}^{+} + \nu  \textbf{\emph{s}} +\textbf{\emph{a}}\right) \sim S_k(\textbf{w}), \quad \quad |t|  \gg 1,
\]
where
\[  \label{wdef}
\textbf{w}=\left(x_1^+, \hspace{0.04cm} T_2, \hspace{0.04cm} 0,  \hspace{0.04cm} 0, \hspace{0.04cm} 0, \cdots\right),
\]
and $T_2$ is as given in Eq.~(\ref{T2T3}). By comparing the definition of Schur polynomials $S_k(\textbf{w})$ to the definition of $q_k(z)$ polynomials in Eq.~(\ref{qk}), we see that
\begin{equation} \label{Skorder2}
S_k(\textbf{w})=T_2^{k/2}q_{k}(z),
\end{equation}
where
\[
z=T_2^{-1/2}x_1^+=T_2^{-1/2}\left(\hat{x}+2\textrm{i}  y\right).
\]
Using these formulae and the Laplace expansion of the determinant (\ref{3Nby3Ndet2}) for $\sigma$, we can readily show that the highest $t$-power term of $\sigma$ is
\begin{equation} \label{sigmanmax2}
\sigma \sim |\mu_0|^2 \hspace{0.05cm} |T_2|^{\rho} \left|W_{\Lambda}(z)\right|^2,
\quad \quad |t| \gg 1,
\end{equation}
where $\rho$ is given in Eq.~(\ref{defrho}), and $\mu_0=2^{-N(N-1)/2}$. Inserting this leading-order term of $\sigma$ into Eq.~(\ref{Schpolysolu}), we see that the solution $u_{\Lambda} (x,y,t)$ approaches zero when $|t|\to \infty$,  except at or near $(\hat{x}, y)$ locations $\left(\hat{x}_0, \hat{y}_{0}\right)$,
i.e., at or near $(x, y)$ locations $\left(12 t+\hat{x}_0, \hat{y}_{0}\right)$, where
\[  \label{z0def2}
z_0=T_2^{-1/2}\left(\hat{x}_0 +2\textrm{i}  \hat{y}_0\right)
\]
is a root of the Wronskian-Hermit polynomial $W_{\Lambda}(z)$. Solving this equation, we get
\[\label{x0t03}
\hat{x}_0=\Re\left[z_0 T_2^{1/2}\right], \quad
\hat{y}_0=\frac{\Im\left[z_0 T_2^{1/2}\right]}{2 },
\]
which are the leading-order terms of $(x_0, y_0)$ in Eq. (\ref{x0t02}) of Theorem 3. Due to our requirement of $\sqrt{\hat{x}^2+y^2}=O(|t|^{1/2})$, $z_0$ in the above equation should be nonzero.

To derive the solution behavior near this $(x, y)=\left(12 t+\hat{x}_0, \hat{y}_{0}\right)$ location, we perform a more refined asymptotic analysis. Our starting point is a more accurate asymptotics for $S_{k}\left(\textbf{\emph{x}}^{+} + \nu \textbf{\emph{s}} +\textbf{\emph{a}}\right)$,
\[ \label{Skxnasym2}
S_k(\textbf{\emph{x}}^{+} + \nu \textbf{\emph{s}}+\textbf{\emph{a}})=S_k(\hat{\textbf{w}}) \left[1+O\left(|t|^{-1}\right)\right], \quad \quad |t| \gg 1,
\]
where
\[
\hat{\textbf{w}}=\left(x_1^+, \hspace{0.04cm} x_2^+, \hspace{0.04cm} T_3,  \hspace{0.04cm} 0, \hspace{0.04cm} 0, \hspace{0.04cm} 0, \cdots\right)=\textbf{w}+\left(0, \hat{x}_2^+, T_3, \hspace{0.04cm} 0, \hspace{0.04cm} 0, \hspace{0.04cm} 0, \cdots\right),
\]
$\textbf{w}$ is given in (\ref{wdef}), and
\[
\hat{x}_2^+\equiv \frac{1}{2}\hat{x}+2\textrm{i} y.
\]
The asymptotics (\ref{Skxnasym2}) holds since $a_1=s_1=0$. From the definition (\ref{Elemgenefunc}) of Schur polynomials and the above equation, we can relate  $S_k(\hat{\textbf{w}})$ and $S_k(\textbf{w})$ as
\[
S_k(\hat{\textbf{w}})=\sum_{j=0}^k b_j S_{k-j}(\textbf{w}),
\]
where $b_j$ are the coefficients in the expansion
\[
e^{\hat{x}_2^+ \epsilon^2+T_3 \hspace{0.02cm} \epsilon^3}=\sum_{j=0}^\infty b_j \epsilon^j.
\]
Notice that $b_0=1$, $b_1=0$, $b_2=\hat{x}_2^+$, and $b_3=T_3$. In addition, $(\hat{x}, y)=O(|t|^{1/2})$ from Eq.~(\ref{x0t03}), and $S_k(\textbf{w})=O(|t|^{k/2})$ in view of Eq.~(\ref{Skorder2}). Utilizing these relations, we find that
\[ \label{Skxnasym2b}
S_k(\textbf{\emph{x}}^{+} + \nu \textbf{\emph{s}}+\textbf{\emph{a}})=\left[ S_k(\textbf{w})+ \hat{x}_2^+ S_{k-2}(\textbf{w})+T_3 \hspace{0.02cm} S_{k-3}(\textbf{w}) \right] \left[1+O\left(|t|^{-1}\right)\right], \quad \quad |t| \gg 1.
\]

With this formula (\ref{Skxnasym2b}), we can now determine the asymptotic expression of $\sigma$ in Eq.~(\ref{3Nby3Ndet2}) in the neighborhood of $(x, y)=\left(12 t+\hat{x}_0, \hat{y}_{0}\right)$ at large $t$. The Laplace expansion of this determinant is very similar to Eq.~(\ref{sigmanLap}) of the previous subsection. Using this Laplace expansion and similar techniques as in Refs.~\cite{YangYang21a,YangYang21b}, we can readily find that
\begin{eqnarray} \label{sigma9}
\sigma(x,y,t) = \left|\mu_0\right|^2 \hspace{0.06cm} \left|W_{\Lambda}'(z_0)\right|^2 |T_2|^{\rho-1}
\left( \left| (x-12 t-\hat{x}_0)+ 2 \textrm{i}  (y-\hat{y}_0) + \Delta \right|^2+ \frac{1}{4}
\right)\left[1+O\left(|t|^{-1/2}\right)\right],
\end{eqnarray}
where $\mu_0$ is given below Eq.~(\ref{sigmanmax2}), and $\Delta=\Delta(\Lambda, z_0)$ is an $O(1)$ complex constant given by
\begin{eqnarray}
&& \Delta=\frac{1}{W_{\Lambda}'(z_0)}\left\{\lambda \sum_{j=1}^{N} \det_{1 \leq i \leq N} \left[ q_{n_i}, \cdots ,   q_{n_i-(j-2)}, \hspace{0.05cm} q_{n_i-(j-1)-2}, \hspace{0.05cm} q_{n_i-j}, \cdots, q_{n_i-(N-1)} \right]_{z=z_0}  \right.   \nonumber \\
&& \hspace{1.8cm} \left. + \frac{4}{3} \sum_{j=1}^{N} \det_{1 \leq i \leq N} \left[
q_{n_i}, \cdots ,   q_{n_i-(j-2)}, \hspace{0.05cm} q_{n_i-(j-1)-3}, \hspace{0.05cm} q_{n_i-j}, \cdots, q_{n_i-(N-1)}\right]_{z=z_0} \right\},   \label{Delta}
\end{eqnarray}
and
\[
\lambda=\left\{  \begin{array}{ll} \frac{1}{2}\Re(z_0)+\textrm{i} \Im(z_0), & \mbox{when} \hspace{0.1cm} t<0, \\
\Re(z_0)+\frac{1}{2}\textrm{i}\Im(z_0), & \mbox{when} \hspace{0.1cm} t>0. \end{array} \right.
\]
The former determinant in Eq.~(\ref{Delta}) is the Wronskian-Hermit determinant in Eq.~(\ref{defWH}) but with the $j$-th column $\{q_{n_i-(j-1)}\}$ replaced by $\{q_{n_i-(j-1)-2}\}$, i.e., reducing the subindex value of this column by two, while the latter determinant in (\ref{Delta}) is the Wronskian-Hermit determinant (\ref{defWH}) with the $j$-th column replaced by $\{q_{n_i-(j-1)-3}\}$, i.e., reducing its subindex value by three.

The complex constant $\Delta$ in Eq.~(\ref{sigma9}) can be absorbed into $(\hat{x}_0, \hat{y}_0)$. After this absorption and rearranging terms, Eq.~(\ref{sigma9}) becomes
\begin{eqnarray} \label{sigma99}
\sigma(x,y,t) = \left|\mu_0\right|^2 \hspace{0.06cm} \left|W_{\Lambda}'(z_0)\right|^2 |T_2|^{\rho-1}
\left[ \left(x-12  t -x_0\right)^2+4  (y-y_0)^2+ \frac{1}{4}\right]\left[1+O\left(|t|^{-1/2}\right)\right],
\end{eqnarray}
where
\[
x_0= \Re\left[z_0 (-12t)^{1/2}\right] -\Re(\Delta), \quad
y_0=\frac{\Im\left[z_0 (-12t)^{1/2}\right]}{2} -\frac{\Im(\Delta)}{2 }.
\]
These $(x_0, y_0)$ formulae contain the explicit $O(1)$ corrections to the leading $O(|t|^{1/2})$ terms, and are thus more
complete than Eq.~(\ref{x0t02}) in Theorem 3. Substituting the above $\sigma$ asymptotics (\ref{sigma99}) into Eq.~(\ref{Schpolysolu}), the asymptotics (\ref{x0t02})-(\ref{Theorem3asym}) for the outer region of Theorem 3 are then proved.

\subsubsection{Proof for the inner region}

In the inner region, where $\sqrt{\hat{x}^2+y^2}\le O(|t|^{1/3})$, a separate asymptotic analysis is needed,
because the previous $S_k$ asymptotics (\ref{Skxnasym1}) and (\ref{Skxnasym2}) do not hold. In this inner region, our analysis needs to split into two cases, depending on whether $\sqrt{\hat{x}^2+y^2}$ is $O(|t|^{1/3})$ or $O(1)$.

When $\sqrt{\hat{x}^2+y^2}=O(|t|^{1/3})$, it is easy to see from the Laplace expansion of the $\sigma$ determinant (\ref{3Nby3Ndet2}) that, at large $|t|$, the highest $t$-power term of $\sigma$ comes from the index choices of $\nu_{j}=j-1$, i.e.,
\begin{eqnarray} \label{sigmanLap5}
&& \hspace{-0.8cm} \sigma\sim \left| \det_{1 \leq i, j\leq N} \left[\frac{1}{2^{j-1}} S_{n_i+1-j}(\textbf{\emph{x}}^{+} + (j-1) \textbf{\emph{s}} +\textbf{\emph{a}}) \right] \right|^2, \qquad |t|\gg 1.
\end{eqnarray}
For the determinant in the above equation, we reorganize its rows by grouping odd-$n_i$ rows together (in ascending order of $n_i$), followed by even-$n_i$ rows (also in ascending order of $n_i$). We also rewrite $S_k(\textbf{\emph{x}}^{+} + \nu \textbf{\emph{s}}+\textbf{\emph{a}})$ as
\[ \label{polyrelation}
S_k(\textbf{\emph{x}}^{+} + \nu \textbf{\emph{s}}+\textbf{\emph{a}}) = \sum_{j=0}^{\left[k/2\right]} \frac{T_2^{j}}{j!} S_{k-2j}(\textbf{h}+ \nu \textbf{\emph{s}}+\textbf{\emph{a}}),
\]
where
\[
\textbf{h}\equiv \textbf{\emph{x}}^{+}-(0, T_2, 0, 0, \cdots)=(x_1^+, \hat{x}_2^+, x_3^+, x_4^+, \cdots),
\]
and $[a]$ represents the largest integer less than or equal to $a$. In addition, we notice that when $\sqrt{\hat{x}^2+y^2}=O(|t|^{1/3})$,
\[
S_{k}(\textbf{h}+ \nu \textbf{\emph{s}}+\textbf{\emph{a}}) \sim S_{k}(\hat{\textbf{h}}),
\]
where
\[
\hat{\textbf{h}}=(x_1^+, 0, T_3, 0, 0,  \cdots),
\]
and
\[
S_{k}(\hat{\textbf{h}})=\left(-3T_3/4\right)^{k/3} p_k(z),
\]
where $z$ is as given in Eq.~(\ref{defz1}). Inserting these formulae into the above reorganized determinant and performing row operations to eliminate certain high powers of $T_2$ in lower rows of the odd-$n_i$-index group as well as the even-$n_i$-index group, we find that the highest $t$-power term of $\sigma$ from Eq.~(\ref{sigmanLap5}) is
\[
\sigma \sim \gamma_0 \hspace{0.05cm} |T_2|^{N_W}|3T_3/4|^{\frac{\hat{d}(\hat{d}+1)}{3}} \left|H(z) \right|^2,
\]
where $\gamma_0$ is a certain positive constant, $N_W$ is given in Eqs.~(\ref{defNW})-(\ref{defNW2}), $\hat{d}$ is defined in Eq.~(\ref{defdhat}), $H(z)$ is the determinant
\[
H(z)=\det \left[\begin{array}{cccc}
p_1(z) & p_0(z) & p_{-1}(z) & \cdots \\
p_3(z) & p_2(z) & p_1(z) & \cdots \\
\vdots & \vdots &   \vdots & \vdots \\
p_{2k_{odd}-1}(z) &   p_{2k_{odd}-2}(z) & p_{2k_{odd}-3}(z) & \cdots \\
p_0(z) & p_{-1}(z) & p_{-2}(z) & \cdots           \\
p_2(z) & p_1(z) & p_0(z) & \cdots \\
\vdots & \vdots &   \vdots & \vdots \\
p_{2k_{even}-2}(z) &   p_{2k_{even}-3}(z) & p_{2k_{even}-4}(z) & \cdots  \end{array} \right],
\]
and $k_{odd}$, $k_{even}$ are the numbers of odd and even elements in the index vector $(n_1, n_2, \dots, n_N)$ respectively. Clearly, this $H$ determinant can be reduced to
\[
H(z)=Q_{\hat{d}}(z).
\]
Thus, when $\sqrt{\hat{x}^2+y^2}=O(|t|^{1/3})$,
\[
\sigma \sim \gamma_0 \hspace{0.05cm} |T_2|^{N_W}|3T_3/4|^{\frac{\hat{d}(\hat{d}+1)}{3}} \left|Q_{\hat{d}}(z) \right|^2.
\]
In view of Eq.~(\ref{Schpolysolu}), this asymptotics shows that the solution $u_\Lambda(x,y,t)$ is asymptotically zero in this region,
except when $\hat{d}>0$ (i.e., zero is a root of the Wronskian-Hermit polynomial $W_\Lambda(z)$), and when $(x, y)$ is at or near the location $\left(12 t+\hat{x}_0, \hat{y}_{0}\right)$, where
\[
z_0=(-3T_3/4)^{-1/3}\left(\hat{x}_0 +2\textrm{i}  \hat{y}_0\right)
\]
is a root of the Yablonskii--Vorob'ev polynomial $Q_{\hat{d}}(z)$. Solving this equation, we get $(\hat{x}_0, \hat{y}_0)$ values that are the leading-order terms in Eq.~(\ref{x0t09}) of Theorem 3. Since $\sqrt{\hat{x}^2+y^2}=O(|t|^{1/3})$, root $z_0$ in the above equation should be nonzero.

We can further show that, near this $(x, y)=\left(12 t+\hat{x}_0, \hat{y}_{0}\right)$ location lies a fundamental lump. This calculation is similar to that we did in the proof of Theorem 2 and the earlier part of this proof of Theorem 3. Specifically, we can show that in the $O(1)$ neighborhood of this location,
\begin{eqnarray} \label{sigmalast}
\sigma(x,y,t) = \gamma_0 \hspace{0.05cm} |T_2|^{N_W}|3T_3/4|^{\frac{\hat{d}(\hat{d}+1)-2}{3}}
\left|Q_{\hat{d}}'(z_0)\right|^2
\left( \left| (x-12 t-\hat{x}_0)+ 2 \textrm{i}  (y-\hat{y}_0) + \hat{\Delta} \right|^2+ \frac{1}{4}
\right)\left[1+O\left(|t|^{-1/3}\right)\right],
\end{eqnarray}
where $\hat{\Delta}=\hat{\Delta}(\Lambda, z_0)$ is an $O(1)$ constant. This $\hat{\Delta}$ is the analog of a similar quantity $\Delta$ which we derived in Eq.~(\ref{Delta}) for a fundamental lump in the outer region.
It is easy to see that the above $\sigma(x,y,t)$ gives a fundamental lump, whose position is at $(x, y)=\left(12 t+x_0, y_{0}\right)$, where
\[ \label{x0y011}
x_0= \Re(z_0) \hspace{0.05cm} (12t)^{1/3}-\Re(\hat{\Delta}), \quad
y_0=\frac{\Im(z_0)}{2}(12t)^{1/3}-\frac{1}{2}\Im(\hat{\Delta}),
\]
which matches (\ref{x0t09}) in Theorem 3. In addition, the error of this prediction is $O\left(|t|^{-1/3}\right)$.

In the center region where $\hat{x}^2+y^2=O(1)$, we can use the technique of Appendix C in Ref.~\cite{YangYang21a} to show that at large time, if zero is a root of the Yablonskii--Vorob'ev polynomial $Q_{\hat{d}}(z)$, i.e., if $\hat{d}\equiv 1 \hspace{0.1cm} \mbox{mod} \hspace{0.1cm} 3$, then $u_\Lambda(x,y,t)$ would approach a fundamental lump located in the $O(1)$ neighborhood of the wave center $(\hat{x}, y)=(0,0)$. If zero is not a root of $Q_{\hat{d}}(z)$, then $u_\Lambda(x,y,t)$ would approach zero in this center region as $|t|\to \infty$. Details will be omitted for brevity. It may be more illuminating for us to point out that, the leading-order term of the previous asymptotic formula (\ref{sigmalast}), which was derived for the region of $\sqrt{\hat{x}^2+y^2}=O(|t|^{1/3})$ and nonzero roots $z_0$ of $Q_{\hat{d}}(z)$, turns out to be valid for the $\hat{x}^2+y^2=O(1)$ region and the zero root $z_0$ of $Q_{\hat{d}}(z)$ as well [except for the relative error term, which is now $O(|t|^{-1})$ rather than $O(|t|^{-1/3})$]. In other words, if zero is a root of $Q_{\hat{d}}(z)$, then setting $z_0=0$ in the leading term of (\ref{sigmalast}), we would get the correct asymptotic fundamental lump in the $\hat{x}^2+y^2=O(1)$ region. In particular, the location of this fundamental lump would be at $(\hat{x}, y)=(x_0, y_0)$, i.e.,
$(x, y)=(12t+x_0, y_0)$, where $x_0$ and $y_0$ are given by (\ref{x0y011}) with $z_0=0$ and $\hat{\Delta}=\hat{\Delta}|_{z_0=0}$. This completes the proof of Theorem~3.

\section{Summary and Discussion}

In this article, we have analytically studied pattern formation in higher-order lumps of the KP-I equation at large time.
For a broad class of these higher-order lumps, we have shown that two types of solution patterns appear at large time. The first type of patterns comprise fundamental lumps arranged in triangular shapes, which are described analytically by root structures of Yablonskii--Vorob'ev polynomials. As time evolves from large negative to large positive, this triangular pattern reverses its $x$-direction. The second type of solution patterns comprise fundamental lumps arranged in non-triangular shapes in the outer region, which are described analytically by nonzero-root structures of Wronskian--Hermit polynomials, together with possible fundamental lumps arranged in triangular shapes in the inner region, which are described analytically by root structures of  Yablonskii--Vorob'ev polynomials. When time evolves from large negative to large positive, the non-triangular pattern in the outer region switches its $x$ and $y$ directions, while the triangular pattern in the inner region reverses its $x$-direction. We have also compared these predicted patterns with true solutions, and excellent agreement is observed.

In this pattern analysis of higher-order lumps, we have set the spectral parameter $p=1$ without any loss of generality (see Remark 4 of Sec.~2). Because of this, lump patterns we have predicted at large time are all $y$-symmetric (see Figs. 3, 5 and 7), since root structures of Yablonskii--Vorob'ev and Wronskian-Hermit polynomials are symmetric with respect to the real-$z$ axis. However, under the Galilean transformation (\ref{Galilean}), these $y$-symmetric lump patterns can become skewed and $y$-asymmetric, and these $y$-asymmetric patterns correspond to complex spectral parameters $p$. Thus, $y$-asymmetric lump patterns also exist in the KP-I equation, and such patterns can be obtained from the $y$-symmetric ones through the Galilean transformation.

Are there other patterns of higher-order lumps at large time? The answer is yes. Notice that in this article, we have assumed internal-parameter vectors $\textbf{\emph{a}}_{i}$ of higher-order lumps to be equal to each other [see Eq.~(\ref{cond_a})]. If these parameter vectors are allowed to differ from each other, then the analytical results at large time will become different. This problem will not be pursued in this paper, and will be left for future studies.

In a very recent preprint \cite{Ling2021}, the authors also derived higher-order lumps in the KP-I equation and studied their large-time patterns through Darboux transformation, and showed that their large-time patterns are described analytically by root structures of Yablonskii--Vorob'ev polynomials. Obviously, the higher-order lump solutions they derived are a very special class of solutions which correspond to the index vector $\Lambda=(1, 3, 5, \dots, 2N-1)$ and under $\textbf{\emph{a}}_{i}$ parameter constraints (\ref{cond_a}) in our general solutions of Theorem 1, and their large-time pattern results are largely equivalent to our Theorem 2. However, their error estimate of $O(|t|^{-2/3})$ for fundamental-lump predictions far away from the wave center is different from our $O(|t|^{-1/3})$ in Theorem 2, and we have verified numerically that their error estimate is incorrect. More importantly, those authors have not considered the more general higher-order lumps corresponding to the index vector $\Lambda\ne (1, 3, 5, \dots, 2N-1)$ in our Theorem 1, nor their large-time solution patterns. These latter patterns are the contents of our Theorem 3 (see also our Figs.~5-8).

\section*{Acknowledgment}
This material is based upon work supported by the National Science Foundation under award number DMS-1910282, and the Air Force Office of Scientific Research under award number FA9550-18-1-0098.

\begin{center}
\textbf{Appendix}
\end{center}

In this appendix, we briefly derive the bilinear higher-order lump solutions presented in Theorem 1.

Under the variable transformation $u=2(\log\tau)_{xx}$ and notations of $x_1=x, x_2=\textrm{i} y$ and $x_3=-4t$, the KP-I equation (\ref{KPI}) is converted to the bilinear equation
\begin{equation} \label{KPIbi}
(D_{x_1}^4-4D_{x_1}D_{x_3}+3D_{x_2}^2) \, \tau \cdot \tau=0,
\end{equation}
where $D$ is Hirota bilinear differential operator. It is well known that if $m_{ij}$, $\phi_i$ and $\psi_j$ are functions of $(x_1, x_2, x_3)$ and satisfy the following differential equations
\begin{eqnarray}
&& \partial_{x_1}m_{ij}=\phi_i \psi_j,   \label{DE1} \\
&& \partial_{x_n}\phi_i  =  \partial_{x_1}^n \phi_i,  \hspace{1.5cm} n=2, 3,   \label{DE2}\\
&& \partial_{x_n}\psi_j  =  (-1)^{n-1}\partial_{x_1}^n \psi_j, \quad n=2, 3,   \label{DE3}
\end{eqnarray}
then the $\tau$ function
\begin{equation}   \label{tau}
\tau=\det_{1\le i,j\le N}\left(m_{ij}\right)
\end{equation}
would satisfy the above bilinear equation \cite{Hirota_book}. To derive higher-order lump solutions, we define $m_{ij}$, $\phi_i$ and $\psi_j$ as
\[  \label{mijAppend}
m_{ij}=\mathcal{A}_i \mathcal{B}_{j} \frac{1}{p+q} e^{\xi_i+\eta_j}, \quad \phi_i=\mathcal{A}_i e^{\xi_i}, \quad
\psi_j=\mathcal{B}_{j} e^{\eta_j},
\]
where
\begin{eqnarray}  \label{AiBj}
\mathcal{A}_{i}=\frac{1}{ n_i !}(p\partial_{p})^{n_i}, \quad
\mathcal{B}_{j}=\frac{1}{ n_j !}(q\partial_{q})^{n_j},
\end{eqnarray}
\[
\xi_i=px_1+p^2x_2+p^3x_3+\xi_{i,0}(p), \quad \eta_j=qx_1-q^2x_2+q^3x_3+\eta_{j,0}(q),
\]
$(n_1, n_2, \cdots, n_N)$ is a vector of arbitrary positive integers, $p, q$ are arbitrary complex constants, and $\xi_{i,0}(p)$, $\eta_{j,0}(q)$ are arbitrary complex functions of $p$ and $q$. It is easy to see that these $m_{ij}$, $\phi_i$ and $\psi_j$ functions satisfy the differential equations (\ref{DE1})-(\ref{DE3}). Thus, the above $\tau$ function would satisfy the bilinear equation (\ref{KPIbi}). To guarantee that this $\tau$ function is real-valued, we impose the parameter constraints
\[
q=p^*, \quad \eta_{j,0}(q)=[\xi_{j,0}(p)]^*.
\]
Under these constraints, $\eta_j=\xi_j^*$, $m_{n_i, n_j}^*=m_{n_j, n_i}$, and thus $\tau$ in (\ref{tau}) is real. Following the technique of Ref.~\cite{OhtaJY2012}, we can further show this $\tau$ is positive. Thus, the resulting function $u=2(\log\tau)_{xx}$ is a real-valued solution to the KP-I equation~(\ref{KPI}).

Next, we need to simplify the matrix elements of this $\tau$ determinant and derive their more explicit algebraic expressions. This simplification is very similar to that we performed in \cite{OhtaJY2012,YangYang3wave}. By expanding $\xi_{i,0}(p)$ into a certain series containing complex parameters $\textbf{\emph{a}}_{i}=\left( a_{i,1}, a_{i,2}, \cdots\right)$ and repeating the calculations of \cite{OhtaJY2012,YangYang3wave}, we can show that the matrix element $m_{ij}$ in (\ref{mijAppend}) can be reduced to the expression given in Eq.~(\ref{Schmatrimnij}) of Theorem 1.

We would like to make a comment here regarding the choice of differential operators in Eq.~(\ref{AiBj}). Obviously, we can also choose more general forms of these differential operators, such as
\begin{eqnarray}  \label{AiBj2}
\mathcal{A}_{i}=\frac{1}{ n_i !}\left[f(p)\partial_{p}\right]^{n_i}, \quad
\mathcal{B}_{j}=\frac{1}{ n_j !}\left[f(q)\partial_{q}\right]^{n_j},
\end{eqnarray}
where $f(p)$ is an arbitrary function, and the resulting $\tau$ function (\ref{tau}) would still satisfy the bilinear equation (\ref{KPIbi}). However, such additional freedoms in the differential operators will not produce new higher-order lump solutions. To see why, we can rewrite this $\mathcal{A}_{i}$ as
\[
\mathcal{A}_{i}=\frac{1}{ n_i !}\left[\frac{f(p)}{p} p\partial_{p}\right]^{n_i}=\sum_{k=0}^{n_i} c_{i,k}\frac{1}{(n_i-k)!} (p\partial_p)^{n_i-k},
\]
where $c_{i,k}$ are $p$-dependent complex constants. Similar treatments can be made on $\mathcal{B}_{j}$. These differential operators in summation form are similar to those taken in Ref.~\cite{OhtaJY2012}. We can directly show that the $m_{ij}$ matrix element with these differential operators of summation form can be converted to one with these differential operators as a single term in (\ref{AiBj}), after parameters $\textbf{\emph{a}}_{i}$ in the series expansion of $\xi_{j,0}(p)$ are redefined properly. Thus, no new solutions are produced.

\end{document}